\documentclass{aa}
\usepackage{graphicx,natbib}
\bibpunct{(}{)}{;}{a}{}{,} 
\begin{document}
   \title{An H$\alpha$ survey aiming at the detection of extraplanar diffuse 
ionized gas in halos of edge--on spiral galaxies
\thanks{Based on observations collected at the European Southern Observatory, 
Chile (ESO No.~63.N--0070, ESO No.~64.N--0034, ESO No.~65.N.--0002).
Figures 22-54 are only available in electronic form at 
http://www.edpsciences.org}}

   \subtitle{II. The H$\alpha$ survey atlas and catalog}

   \author{J. Rossa \thanks{Visiting Astronomer, German--Spanish Astronomical 
Centre, Calar Alto, operated by the Max--Planck--Institute for Astronomy, 
Heidelberg, jointly with the Spanish National Commission for Astronomy.}
          \inst{1,2}
          \and
          R.--J. Dettmar\inst{1}
          }

   \offprints{J. Rossa}

   \institute{Astronomisches Institut, Ruhr--Universit\"at Bochum,
              D--44780 Bochum, Germany\\
              \email{jrossa@stsci.edu, 
              dettmar@astro.ruhr-uni-bochum.de}
         \and
             Space Telescope Science Institute, 3700 San Martin Drive,  
             Baltimore, MD 21218, U.S.A. (present address)}

   \date{Received 14 February 2003 / Accepted 6 May 2003}

   \abstract{
In this second paper on the investigation of extraplanar diffuse ionized gas 
in nearby edge--on spiral galaxies we present the actual results of the 
individual galaxies of our H$\alpha$ imaging survey. A grand total of 74 
galaxies have been studied, including the 9 galaxies of a recently studied 
sub--sample \citep{Ro00}. 40.5\% of all studied galaxies reveal extraplanar 
diffuse ionized gas, whereas in 59.5\% of the survey galaxies no extraplanar 
diffuse ionized gas could be detected. The average distances of this extended 
emission above the galactic midplane range from 1--2\,kpc, while individual 
filaments in a few galaxies reach distances of up to $|z|\sim6$\,kpc. In 
several cases a pervasive layer of ionized gas was detected, similar to the 
Reynolds layer in our Milky Way, while other galaxies reveal only extended 
emission locally. The morphology of the diffuse ionized gas is discussed for 
each galaxy and is compared with observations of other important ISM 
constituents in the context of the disk--halo connection, in those cases 
where published results were available. Furthermore, we present the 
distribution of extraplanar dust in these galaxies, based on an analysis of 
the unsharp--masked R--band images. The results are compared with the 
distribution of the diffuse ionized gas.  
          \keywords{galaxies: halos --
             galaxies: spiral --
             galaxies: starburst --
             galaxies: ISM --
             galaxies: structure
               }
   }
\titlerunning{An H$\alpha$ survey aiming at the detection of eDIG in edge--on 
spiral galaxies}
   \maketitle
%
%__________________________________________________________________

\section{Introduction}

Our current knowledge of the morphology of the extraplanar diffuse ionized 
gas (eDIG) in normal (non--starburst) edge--on spiral galaxies rests on a 
few investigations of small galaxy samples \citep{Pi94,Ra96,Hoo99,Ro00}, as 
well as on some additional studies of individual galaxies 
\citep[e.g.,\,][]{De90,Ra92,Fe96,Dom97}. There have been a few galaxies 
studied during the last decade, however, the question was raised, whether 
or not the presence of eDIG is a common phenomenon among all types of 
galaxies, or whether it is indeed a direct consequence of the strength of 
the star formation activity, both on local and global scales 
\citep{Ra96,Ro00}. To answer this question a systematic investigation is 
obligatory. We have therefore conducted, for the first time, a large survey 
of nearby {\em non--starburst} edge--on spiral galaxies, aiming at a 
quantitative morphological study of gaseous halos, based on the broad 
coverage of the strength of SF activity in the underlying galaxy disks (i.e. 
broad coverage of $L_{\rm FIR}$) of these galaxies. We present the observed 
morphological results, based on the H$\alpha$ imaging observations, and 
describe the derived DIG characteristics for each galaxy in this paper in 
greater detail. For information on the scientific background on eDIG and its 
detection in external galaxies we refer the reader to our previous works 
\citep{Ro00,RoDe01}, and references therein and to some older review articles 
\citep{De92,Da97} for a more detailed overview on this topic. 
  
The current paper is structured as follows. In Sect.~2 we present some 
details on the observations and data reduction procedures. In Sect.~3 we show 
the actual results for the galaxies, while in Sect.~4 we discuss the 
extraplanar dust. Then the atlas is presented (available only 
electronically at EDP Sciences), where we display two galaxies on each page. 
Each of the two columns per page consists of the R--band (top), the 
unsharp--masked R--band (middle), and the continuum subtracted H$\alpha$ 
image (bottom). Finally, in Sect.~5 we summarize briefly the results. 

%_______________________________________________________________________

\section{Observations and data reduction}

\subsection{H$\alpha$ observations}

The edge--on spiral galaxies, which form the basis of this H$\alpha$ survey, 
have been observed in five individual observing runs between 1999 and 2000. 
The northern hemisphere objects (Runs NH--1, NH--2) have been observed with 
the CAHA\footnote{Centro Astron\'omico Hispano Alem\'an} 2.2\,m telescope at 
the Calar Alto Observatory in southern Spain, and the southern hemisphere 
objects (Runs SH--1, SH--2, SH--3) have been observed with the Danish 1.54\,m 
telescope at the European Southern Observatory at La Silla/Chile (see 
Table~1 for a summary of the individual observing runs). Six additional 
targets, that were selected on the basis of our selection criteria, have 
also been included into this survey. These were galaxies, already observed 
several years ago in past observing runs, or that were at our disposal, so 
we did not have to re--observe them (Runs SH--a, SH--b, SH--c). The basic 
galaxy properties, including the coordinates, galaxy types, distances, radial 
velocities, sizes, inclinations, and magnitudes, are listed in Table~2.

\begin{table}[h]
\caption[]{Individual observing runs for the new H$\alpha$ survey}
\begin{flushleft}
\begin{center}
\begin{tabular}{ccc}
\noalign{\smallskip}
\hline\hline
Run No. & Dates & Telescope \\
\hline
NH--1 & 19--22/02/1999 & CAHA2.2 \\
NH--2 & 08--13/08/1999 & CAHA2.2 \\
SH--1 & 07--10/07/1999 & DAN1.54 \\
SH--2 & 09--13/11/1999 & DAN1.54 \\
SH--3 & 29/07--01/08/2000 & DAN1.54 \\
SH--a & 22/02/1993 & ESO2.2 \\
SH--b & 07--08/05/1991 & NTT \\
SH--c & 10/02/1995 & NTT \\ 
%\noalign{\smallskip}
\hline
\end{tabular}
\end{center}
\end{flushleft}    
\end{table}

The observations for the northern hemisphere galaxies have been performed 
with CAFOS\footnote{Calar Alto Faint Object Spectrograph} in direct imaging 
mode, attached to the CAHA 2.2\,m telescope. The used CCD chip was a 2048 
$\times$ 2048 pixel SITe\#1d, with a pixel size of 24$\mu$m. The pixel scale 
was $0\farcs53\,\rm{pix}^{-1}$, and the resulting field of view was 
$18\farcm1\times18\farcm1$. The used H$\alpha$ filters were the CA \#658/10 
and the CA \#665/17 filters with a $\Delta\lambda$ = 98\,{\AA} and 
168\,{\AA}, respectively. The used R--band filter (Johnson R) was the CA 
\#641/158 with $\Delta\lambda$ = 1575\,{\AA}. 

The southern hemisphere galaxies have been observed with 
DFOSC\footnote{Danish Faint Object Spectrograph Camera} in imaging mode, 
attached to the Danish 1.54\,m telescope at La Silla/Chile. The used CCD 
camera was a LORAL/LESSER 2\,k $\times$ 2\,k chip with a pixel size of 
15$\mu$m. The pixel scale was $0\farcs39\,\rm{pix}^{-1}$, and the maximum 
field of view (fov) was $13\farcm7\times13\farcm7$. Due to vignetting effects 
at the borders of the filters we have used only the inner 1601 $\times$ 1601 
pixels of the CCD chip in most cases, yielding a fov of 
$10\farcm4\times10\farcm4$. The used H$\alpha$ filters were the ESO \#693, 
\#697, and \#701 filters with a $\Delta\lambda$ = 62.1\,{\AA}, 61.6\,{\AA}, 
and 63.1\,{\AA}, respectively. The broad--band observations, to be used for 
the continuum subtraction, have been acquired through the ESO \#452 (Bessel 
R) filter, which has a $\Delta\lambda=$\,1647\,{\AA}. Instrumental parameters 
for six additional objects can be found elsewhere \citep{Ro00}. The H$\alpha$ 
transmission curves of the filters, which have been used for this survey, 
are shown in Fig.~1. A journal of observations for our galaxies is given in 
Table~3.

\begin{figure}[t]
\begin{center}
\rotatebox{270}{\resizebox{6.0cm}{!}{\includegraphics{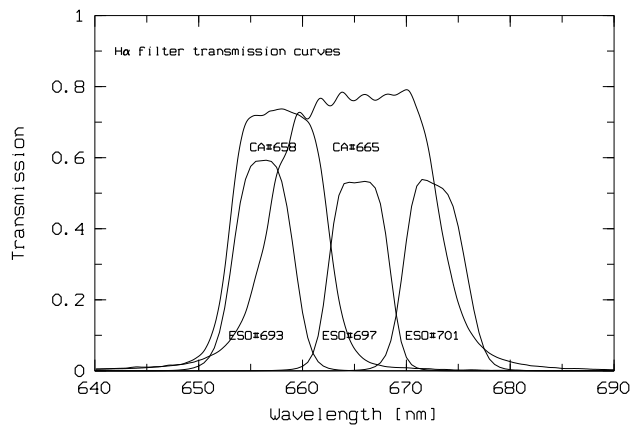}}}
\caption[]{\small Transmission curves for the ESO and Calar Alto H$\alpha$ 
filters (ESO \#693, \#697, \#701 and CA\,\#658/10, CA\,\#665/17), 
respectively. The plot has been generated in the MIDAS environment with the 
use of the implemented ESO filter and edited CAHA filter table--files.}
\end{center}
\end{figure}

\subsection{Data reduction and analysis}

The data reduction has been performed in the usual manner including bias 
level correction, flat--fielding, background correction, registering of the 
stars (alignment of the R--band image with respect to the H$\alpha$ image), 
scaling of the R--band image, and final continuum subtraction. In almost all 
cases the H$\alpha$ images (split into two individual exposures for cosmic 
ray removal) have been combined. All these reduction steps have been 
performed using various IRAF packages. 

For the northern survey the bias correction was performed by using the IRAF 
{\em colbias} task, taking the information of the overscan region into 
account. This was appropriate, since the bias exposures showed diagonal 
stripes. For the southern hemisphere objects the bias correction was done 
using a {\em masterbias} frame (one per night) obtained from median combining 
individual bias frames of each night. The flat--field correction was done 
using a normalized flat--field image created from combining the individual 
flat--field exposures for each filter separately. Final background correction 
was done by subtracting the mean sky countrates from the flat--field 
corrected object frames. These values were estimated from at least three 
regions on each object frame that were neither contaminated by cosmic--ray 
hits nor populated by stars. The R--band frames were aligned and shifted with 
respect to the H$\alpha$ exposures before combining each of the two object 
exposures. Finally the R--band image was scaled to the H$\alpha$ image. The 
scaling factor has been determined from the ratio of the countrates of 
individual foreground stars in the R--band and H$\alpha$ images. Then the 
scaled R--band image was subtracted from the H$\alpha$ image. Due to the 
given bandwidth of the used H$\alpha$ filters also the adjacent 
[\hbox{N\,{\sc ii}}] emission is covered. For the sake of brevity we 
therefore always refer to the emission of H$\alpha$+[\hbox{N\,{\sc ii}}] when 
we speak of an H$\alpha$ image. More details on the continuum subtraction 
are given in Sect.~3.3.

%__________________________________________________________________

\section{Results}

\subsection{General results} 

This current H$\alpha$ survey comprises of 65 nearby edge--on galaxies. 59 
of these galaxies have been observed during the course of this survey, 
whereas we have added 6 additional target galaxies, from which we had 
H$\alpha$ data at our disposal. Taking the 9 edge--on galaxies from our first 
sample \citep{Ro00} into consideration, the survey covers 74 galaxies in 
total. From these 74 galaxies 63 actually are new observations, whereas 11 
galaxies have already been investigated in the DIG context by other 
researchers. We have included them because we wanted some galaxies for an 
intercomparison between our sample and the samples studied by other 
investigators. Furthermore, we wanted to have a more homogeneous sample in 
the sense of FIR luminosity, and thus included also a very few starburst 
galaxies. Needless to say that many starbursts fulfilled our selection 
criteria, but we did not want to re--observe all targets, and were also 
primarily interested in the galaxies with lower SF activity. The description 
of the selection criteria for the H$\alpha$ survey target galaxies are 
presented in detail in Paper~I \citep{RoDe01}. Far below (Table~5) we present 
the DIG morphology of the sample galaxies with information on the vertical 
extent ($|z|$), and the radial extent of the star formation activity 
(R$_{\rm SF}$\,[kpc]).

\begin{table*}
\setcounter{table}{1}
\caption[Basic galaxy parameters]{Basic galaxy parameters}
%\label{T1}
\begin{flushleft}
\begin{minipage}{20cm}\small
\begin{tabular}{llcccccccl}
\noalign{\smallskip}
\hline\hline
Galaxy$^{\footnotesize a}$ & alt. name & R.A. (J2000) & Dec. (J2000) & Type & 
$D$\,[Mpc] & $v_{\rm \tiny \hbox{H\,{\sc i}}}\,[\rm{km\,s^{-1}}]$ & 
$a \times b$ & $i$ & $m_{\rm B}^{\,\,\,\,\,\footnotesize b}$ \\
\hline
\object{NGC\,24} & ESO\,472--16 & $\rm{00^h09^m56.6^s}$ & $-24^\circ57'53''$ 
& Sc & \,~6.8 & \,~554 & $5.8' \times 1.3'$ & 77$^\circ$ & 12.07\\
\object{NGC\,100} & FGC\,0042 & $\rm{00^h24^m02.6^s}$ & $+16^\circ29'09''$ & 
Sc & 11.2 & \,~842 & $5.7' \times 0.6'$ & 84$^\circ$ & 14.00\\
\object{UGC\,260} & FGC\,0046 & $\rm{00^h27^m02.9^s}$ & $+11^\circ35'03''$ & 
Sc & 28.5 & 2135 & $2.7' \times 0.4'$ & 81$^\circ$ & 13.71\\
\object{ESO\,540--16} & FGC\,0082 & $\rm{00^h42^m14.9^s}$ & 
$-18^\circ09'40''$ & SBcd & 20.7 & 1555 & $2.7' \times 0.3'$ & 84$^\circ$ & 
14.32\\
\object{MCG-2-3-16} & FGC\,0090 & $\rm{00^h47^m46.7^s}$ & $-09^\circ53'54''$ 
& SBc & 17.9 & 1345 & $3.0' \times 0.4'$ & 82$^\circ$ & - - - - \\
\object{NGC\,360} & ESO\,79--14& $\rm{01^h02^m51.2^s}$ & $-65^\circ36'34''$ 
& Sc & 30.7 & 2302 & $3.5' \times 0.4'$ & 83$^\circ$ & 13.40\\
\object{NGC\,669} & UGC\,01248 & $\rm{01^h47^m16.1^s}$ & $+35^\circ33'48''$ & 
Sab & 62.4 & 4677 & $3.1' \times 0.6'$ & 79$^\circ$ & 12.97\\
\object{UGC\,1281} & FGC\,0195 & $\rm{01^h49^m32.4^s}$ & $+32^\circ35'32''$ & 
Sc & \,~4.6 & \,~157 & $4.5' \times 0.8'$ & 80$^\circ$ & 12.87\\
\object{NGC\,891} & UGC\,01831 & $\rm{02^h22^m33.1^s}$ & $+42^\circ20'48''$ & 
Sb & \,~9.5 & \,~528 & \hspace{-0.265cm} $14.0' \times 3.0'$ & $88^\circ$ & 
10.84\\
\object{UGC\,2082} & FGC\,0317 & $\rm{02^h36^m16.3^s}$ & $+25^\circ25'29''$ & 
Sc & \,~9.4 & \,~707 & $5.4' \times 0.8'$ & 81$^\circ$ & 13.56\\
\object{IC\,1862} & ESO\,356--15 & $\rm{02^h51^m57.6^s}$ & $-33^\circ20'31''$ 
& Sbc & 85.3 & 6400 & $2.8' \times 0.3'$ & 84$^\circ$ & 14.45\\
\object{NGC\,1247} & FGC\,0396 & $\rm{03^h12^m13.0^s}$ & $-10^\circ28'58''$ & 
Sab & 52.6 & 3945 & $3.4' \times 0.5'$ & 82$^\circ$ & 13.47\\
\object{ESO\,117--19} & PGC\,14337 & $\rm{04^h02^m32.5^s}$ & 
$-62^\circ18'53''$ & SBbc & 71.1 & 5335 & $2.0' \times 0.3'$ & 81$^\circ$ & 
14.58\\
\object{IC\,2058} & ESO\,157--18 & $\rm{04^h17^m54.5^s}$ & $-55^\circ55'57''$ 
& Sc & 18.2 & 1368 & $3.1' \times 0.4'$ & 83$^\circ$ & 13.87\\
\object{ESO\,362--11} & PGC\,017027 & $\rm{05^h16^m39.0^s}$ & 
$-37^\circ06'00''$ & Sbc & 17.9 & 1346 & $4.7' \times 0.8'$ & 80$^\circ$ & 
13.04\\
\object{ESO\,121--6} & FGCE\,0562 & $\rm{06^h07^m29.2^s}$ & 
$-61^\circ48'25''$ & Sc & 16.2 & 1211 & $3.9' \times 0.7'$ & 80$^\circ$ & 
13.43\\
\object{NGC\,2188} & ESO\,364--37 & $\rm{06^h10^m09.5^s}$ & 
$-34^\circ06'22''$ & SBcd & \,~7.9 & \,~749 & $4.8' \times 0.9'$ & 
79$^\circ$ & 12.04\\
\object{ESO\,209--9} & FGCE\,0684 & $\rm{07^h58^m15.6^s}$ & 
$-49^\circ51'22''$ & SBc & 14.9 & 1119 & $6.2' \times 0.9'$ & 82$^\circ$ & 
12.68\\
\object{UGC\,4559} & FGC\,103A & $\rm{08^h44^m08.0^s}$ & $+30^\circ07'07''$ & 
Sb & 27.8 & 2085 & $3.2' \times 0.5'$ & 81$^\circ$ & 14.10\\
\object{NGC\,2654} & PGC\,024784 & $\rm{08^h49^m12.5^s}$ & $+60^\circ13'13''$ 
& SBab & 17.9 & 1342 & $4.2' \times 0.8'$ & 79$^\circ$ & 12.77\\
\object{NGC\,2683} & UGC\,04641 & $\rm{08^h52^m41.0^s}$ & $+33^\circ25'03''$ 
& Sb & \,~5.5 & \,~410 & $8.8' \times 2.6'$ & 77$^\circ$ & 10.36\\
\object{NGC\,3003} & UGC\,5251 & $\rm{09^h48^m36.0^s}$ & $+33^\circ25'18''$ & 
SBc & 19.7 & 1481 & $6.0' \times 1.4'$ & 77$^\circ$ & 12.15\\
\object{NGC\,3221} & UGC\,5601 & $\rm{10^h32^m20.4^s}$ & $+21^\circ34'09''$ & 
SBcd & 54.8 & 4110 & $3.2' \times 0.7'$ & 77$^\circ$ & 13.80\\
\object{NGC\,3365} & UGC\,5878 & $\rm{10^h46^m13.1^s}$ & $+01^\circ48'46''$ & 
Sc & 13.2 & \,~986 & $4.4' \times 0.7'$ & 81$^\circ$ & 13.04\\
\object{NGC\,3501} & UGC\,6116 & $\rm{11^h02^m46.9^s}$ & $+17^\circ59'33''$ & 
Sc & 15.1 & 1133 & $3.5' \times 0.5'$ & 82$^\circ$ & 13.57\\
\object{NGC\,3600} & UGC\,6283 & $\rm{11^h15^m52.1^s}$ & $+41^\circ35'27''$ & 
Sab & \,~9.6 & 1443 & $4.1' \times 0.9'$ & 85$^\circ$ & 12.60\\
\object{NGC\,3628} & UGC\,6350 & $\rm{11^h20^m16.3^s}$ & $+13^\circ35'22''$ & 
Sb & \,~7.7 & \,~847 & \hspace{-0.265cm} $14.8' \times 3.0'$ & 88$^\circ$ & 
10.42\\
\object{NGC\,3877} & UGC\,6745 & $\rm{11^h46^m08.0^s}$ & $+47^\circ29'39''$ & 
Sc & 12.1 & \,~904 & $5.3' \times 1.2'$ & 83$^\circ$ & 11.85\\
\object{NGC\,3936} & ESO\,504--20 & $\rm{11^h52^m20.4^s}$ & 
$-26^\circ54'24''$ & SBbc & 29.0 & 2022 & $3.9' \times 0.6'$ & 81$^\circ$ & 
12.76\\
\object{ESO\,379--6} & FGCE\,0915 & $\rm{11^h53^m03.1^s}$ & 
$-36^\circ38'20''$ & Sbc & 39.8 & 2986 & $2.6' \times 0.4'$ & 81$^\circ$ & 
14.13\\
\object{NGC\,4206} & IC\,3064 & $\rm{12^h15^m16.6^s}$ & $+13^\circ01'30''$ & 
Sc & \,~9.4 & \,~702 & $5.2' \times 1.0'$ & 79$^\circ$ & 12.80\\
\object{NGC\,4216} & UGC\,7284 & $\rm{12^h15^m53.1^s}$ & $+13^\circ08'58''$ & 
SBab & 16.8 & \,~132 & $7.9' \times 1.7'$ & 78$^\circ$ & 10.95\\
\object{NGC\,4235} & IC\,3098 & $\rm{12^h17^m08.9^s}$ & $+07^\circ11'31''$ & 
Sa & 32.1 & 2410 & $3.8' \times 0.8'$ & 78$^\circ$ & 12.64\\
\object{NGC\,4256} & UGC\,7351 & $\rm{12^h18^m44.2^s}$ & $+65^\circ53'58''$ & 
Sb & 33.7 & 2528 & $4.1' \times 0.7'$ & 80$^\circ$ & 12.79\\
\object{NGC\,4388} & UGC\,7520 & $\rm{12^h25^m47.0^s}$ & $+12^\circ39'42''$ & 
Sb & 33.6 & 2518 & $5.5' \times 1.5'$ & 79$^\circ$ & 11.91\\
\object{NGC\,4700} & PGC\,43330 & $\rm{12^h49^m07.8^s}$ & $-11^\circ24'38''$ 
& SBc & 24.0 & 1407 & $3.0' \times 0.6'$ & 78$^\circ$ & 12.71\\
\object{NGC\,4945} & ESO\,219--24 & $\rm{13^h05^m26.2^s}$ & 
$-49^\circ28'15''$ & SBc & \,~7.5 & \,~561 & \hspace{-0.265cm} $20.4' \times 
4.1'$ & 78$^\circ$ & \,~9.15\\
\object{NGC\,5290} & UGC\,8700 & $\rm{13^h45^m19.2^s}$ & $+41^\circ42'55''$ & 
Sbc & 34.4 & 2580 & $3.7' \times 1.0'$ & 79$^\circ$ & 12.93\\
\object{NGC\,5297} & UGC\,8709 & $\rm{13^h46^m24.2^s}$ & $+43^\circ52'25''$ & 
SBbc & 67.5 & 2402 & $5.4' \times 1.2'$ & 77$^\circ$ & 12.37\\
\object{NGC\,5775} & UGC\,9579 & $\rm{14^h53^m57.7^s}$ & $+03^\circ32'40''$ & 
SBc & 26.7 & 1681 & $4.2' \times 1.0'$ & 82$^\circ$ & 12.45\\
\object{ESO\,274--1} & FGCE\,1205 & $\rm{15^h14^m13.6^s}$ & 
$-46^\circ48'45''$ & Sd & 7.0 & \,~522 & \hspace{-0.265cm} $11.0' \times 
1.6'$ & 82$^\circ$ & 12.00\\
\object{NGC\,5965} & UGC\,9914 & $\rm{15^h34^m02.1^s}$ & $+56^\circ41'10''$ & 
Sb & 45.5 & 3412 & $4.7' \times 0.7'$ & 81$^\circ$ & 13.21\\
\object{NGC\,6722} & ESO\,104--33 & $\rm{19^h03^m39.6^s}$ & 
$-64^\circ53'41''$ & Sab & 76.7 & 5749 & $2.9' \times 0.4'$ & 82$^\circ$ & 
13.56\\
\object{IC\,4837A} & ESO\,184--47 & $\rm{19^h15^m15.7^s}$ & 
$-54^\circ07'57''$ & Sab & 38.0 & 2847 & $4.1' \times 0.7'$ & 80$^\circ$ & 
 12.55\\
\object{ESO\,142--19} & PGC\,63351 & $\rm{19^h33^m18.0^s}$ & 
$-58^\circ06'50''$ & S0-a & 56.2 & 4211 & $4.4' \times 1.1'$ & 81$^\circ$ & 
13.56\\
%\noalign{\smallskip}
\hline
\end{tabular}
\end{minipage}
\end{flushleft}
\end{table*}

\begin{table*}
\setcounter{table}{1}
\caption[continued]{continued}
%\label{T1}
\begin{flushleft}
\begin{minipage}{20cm}\small
\begin{tabular}{llcccccccl}
\noalign{\smallskip}
\hline\hline
Galaxy\footnote{All data have been taken from the RC3 \citep{Vau91}, or have 
been calculated from those data,\\ \hspace*{0.53cm}except 
where indicated} & alt. name & 
R.A. (J2000) & Dec. (J2000) & Type & $D$\,[Mpc] & 
$v_{\rm \tiny \hbox{H\,{\sc i}}}\,[\rm{km\,s^{-1}}]$ & $a \times b$ & 
$i$ & $m_{\rm B}$\footnote{taken from NED, and from \citep{ScVi}} \\
\hline
%\noalign{\smallskip}
\object{IC\,4872} & ESO\,142--24 & $\rm{19^h35^m42.3^s}$ & $-57^\circ31'10''$ 
& SBc & 25.7 & 1928 & $3.5' \times 0.4'$ & 83$^\circ$ & 14.03\\
\object{NGC\,6875A} & ESO\,284--24 & $\rm{20^h11^m55.6^s}$ & 
$-46^\circ08'37''$ & SBc & 42.4& 3176 & $2.8' \times 0.5'$ & 80$^\circ$ & 
13.96\\
\object{MCG-1-53-12} & FGC\,2290 & $\rm{20^h49^m52.5^s}$ & 
$-07^\circ01'21''$ & Sc & 79.5 & 5965 & $3.0' \times 0.4'$ & 82$^\circ$ 
& - - - - \\
\object{IC\,5052} & ESO\,74--15 & $\rm{20^h52^m06.3^s}$ & $-69^\circ12'13''$ 
& SBcd & \,~7.9 & \,~591 & $5.9' \times 0.9'$ & 81$^\circ$ & 11.91\\
\object{IC\,5071} & ESO\,47--19 & $\rm{21^h01^m19.7^s}$ & $-72^\circ38'33''$ 
& SABb & 41.6 & 3122 & $3.4' \times 0.8'$ & 76$^\circ$ & 13.28\\
\object{IC\,5096} & ESO\,107--19 & $\rm{21^h18^m21.0^s}$ & 
$-63^\circ45'41''$ & Sbc & 41.9 & 3144 & $3.1' \times 0.5'$ & 81$^\circ$ & 
13.30\\
\object{NGC\,7064} & ESO\,188--9 & $\rm{21^h29^m02.4^s}$ & $-52^\circ45'58''$ 
& SBc & 11.4 & \,~858 & $3.4' \times 0.6'$ & 80$^\circ$ & 12.87\\
\object{NGC\,7090} & ESO\,188--12 & $\rm{21^h36^m28.6^s}$ & 
$-54^\circ33'27''$ & SBc & 11.4 & \,~857 & $7.7' \times 1.4'$ & 80$^\circ$ & 
11.47\\
\object{UGC\,11841} & FGC\,2351 & $\rm{21^h52^m48.0^s}$ & $+38^\circ56'00''$ 
& Scd & 79.9 & 5990 & $2.8' \times 0.2'$ & 86$^\circ$ & - - - - \\
\object{NGC\,7184} & ESO\,601--9 & $\rm{22^h02^m38.5^s}$ & $-20^\circ48'47''$ 
& SBc & 34.1 & 2617 & $6.0' \times 1.5'$ & 76$^\circ$ & 11.91\\
\object{IC\,5171} & ESO\,288--46 & $\rm{22^h10^m56.2^s}$ & $-46^\circ04'54''$ 
& SBb & 34.1 & 2850 & $3.0' \times 0.4'$ & 82$^\circ$ & 13.45\\
\object{IC\,5176} & ESO\,108--20 & $\rm{22^h14^m54.5^s}$ & $-66^\circ50'46''$ 
& SBbc & 23.3 & 1746 & $4.5' \times 0.5'$ & 84$^\circ$ & 13.52\\
\object{NGC\,7339} & UGC\,12122 & $\rm{22^h37^m46.9^s}$ & $+23^\circ47'14''$ 
& SBbc & 17.9 & 1343 & $2.7' \times 0.7'$ & 80$^\circ$ & 13.07\\
\object{NGC\,7361} & IC\,5237 & $\rm{22^h42^m18.0^s}$ & $-30^\circ03'29''$ & 
Sc & 15.9 & 1245 & $3.8' \times 1.0'$ & 80$^\circ$ & 12.90\\
\object{NGC\,7412A} & ESO\,290--28 & $\rm{22^h57^m07.0^s}$ & 
$-42^\circ48'15''$ & SBd & 12.4 & \,~929 & $3.9' \times 0.6'$ & 81$^\circ$ & 
14.30\\
\object{UGC\,12281} & FGC\,2441 & $\rm{22^h59^m12.4^s}$ & $+13^\circ36'21''$ 
& Sc & 34.2 & 2565 & $3.5' \times 0.3'$ & 85$^\circ$ & 14.79\\
\object{NGC\,7462} & ESO\,346--28 & $\rm{23^h02^m46.5^s}$ & 
$-40^\circ50'02''$ & SBbc & 14.1 & 1061 & $4.0' \times 0.9'$ & 84$^\circ$ & 
12.36\\
\object{UGC\,12423} & FGC\,2469 & $\rm{23^h13^m06.0^s}$ & $+06^\circ24'00''$ 
& Sc & 64.5 & 4838 & $3.4' \times 0.4'$ & 83$^\circ$ & 14.41\\
\object{NGC\,7640} & UGC\,12554 & $\rm{23^h22^m06.7^s}$ & $+40^\circ50'43''$ 
& SBc & \,~8.6 & \,~385 & $9.9' \times 2.2'$ & 77$^\circ$ & 11.46\\
\object{ESO\,240--11} & FGCE\,1839 & $\rm{23^h37^m49.4^s}$ & 
$-47^\circ43'35''$ & Sb & 37.9 & 2843 & $5.2' \times 0.5'$ & 84$^\circ$ & 
13.05\\
%\noalign{\smallskip}
\hline
\end{tabular}
\end{minipage}
\end{flushleft}
\end{table*}   

\begin{table*}
\setcounter{table}{2}
\caption[Journal of observations]{Journal of observations}
%\label{T1}
\begin{flushleft}
\begin{minipage}{20cm}\small
\begin{tabular}{lccccccc}
%\noalign{\smallskip}
\hline\hline
Galaxy & Date & Telescope & Instrument & H$\alpha$ filter Id. & 
$t_{\rm exp}$\,(H$\alpha$)\,[s] & $t_{\rm exp}$\,(R)\,[s] & Seeing [$''$]\\
\hline
NGC\,24 & 01/08/2000 & DAN1.54 & DFOSC & ESO\#693 & $2 \times 2700$ & 
$2 \times 600$ & 1.6 \\
NGC\,100 & 10/08/1999 & CAHA2.2 & CAFOS & CA\#658/10 & $ 2\times2700$ & 
$120 + 300$ & \\
UGC\,260 & 12/08/1999 & CAHA2.2 & CAFOS & CA\#658/10 & $ 2\times2700 $ & 
$120+300$ & 1.5 \\
ESO\,540--16 & 01/08/2000 & DAN1.54 & DFOSC & ESO\#693 & $ 2\times2700 $ & 
$2\times600$ & 1.7 \\   
MCG-2-3-16 & 10/11/1999 & DAN1.54 & DFOSC & ESO\#693 & $ 2\times2700 $ & 
$300+600$ & 1.2 \\ 
NGC\,360 & 10/07/1999 & DAN1.54 & DFOSC & ESO\#697 & $ 2\times2700 $ & $600$ 
& \\ 
NGC\,669 & 13/08/1999 & CAHA2.2 & CAFOS & CA\#665/17 & $ 2\times2700 $ &
$120+300$ & 1.7 \\ 
UGC\,1281 & 12/08/1999 & CAHA2.2 & CAFOS & CA\#658/10 & $ 2\times2700 $ & 
$120+300$ & 1.4 \\ 
NGC\,891 & 11/08/1999 & CAHA2.2 & CAFOS & CA\#658/10 & $2\times1800 $ & 
$120 + 300$ & 2.0 \\
UGC\,2082 & 13/08/1999 & CAHA2.2 & CAFOS & CA\#658/10 & $ 1\times1800$ & 
$300$ & \\  
IC\,1862 & 11/11/1999 & DAN1.54 & DFOSC & ESO\#701 & $ 2\times2700 $ & 
$120+600$ & 1.5 \\ 
NGC\,1247 & 10/11/1999 & DAN1.54 & DFOSC & ESO\#697 & $ 2\times2700 $ & 
$120+600$ & 1.6 \\ 
ESO\,117--19 & 22/02/1993 & ESO2.2 & EFOSC2 & ESO\#439 & $ 1\times1800 $ & 
$900$ & \\ 
 & 22/02/1993 & ESO2.2 & EFOSC2 & ESO\#439 & $ 1\times900 $ & -- & \\ 
IC\,2058 & 10/11/1999 & DAN1.54 & DFOSC & ESO\#693 & $ 1\times2700 $ & 
$120+600$ & 1.4 \\ 
 & 10/11/1999 & DAN1.54 & DFOSC & ESO\#693 & $ 1\times2200 $ & -- & \\
ESO\,362--11 & 11/11/1999 & DAN1.54 & DFOSC & ESO\#693 & $ 2\times2700 $ & 
$120+600$ & 1.2 \\ 
ESO\,121--6 & 13/11/1999 & DAN1.54 & DFOSC & ESO\#693 & $ 2\times2700 $ & 
$120+300$ & 1.3 \\ 
NGC\,2188 & 23/02/1993 & ESO2.2 & EFOSC2 & ESO\#694 & $ 2 \times1800 $ &
$600$ & \\ 
ESO\,209--9 & 13/11/1999 & DAN1.54 & DFOSC & ESO\#693 & $ 2\times2700 $ & 
$120+300$ & 1.5 \\ 
UGC\,4559 & 22/02/1999 & CAHA2.2 & CAFOS & CA\#658/10 & $ 2\times1800 $ & 
$720$ & 2.2 \\ 
NGC\,2654 & 21/02/1999 & CAHA2.2 & CAFOS & CA\#658/10 & $ 2\times1800 $ & 
$600$ & \\ 
NGC\,2683 & 19/02/1999 & CAHA2.2 & CAFOS & CA\#658/10 & $ 2\times1800 $ & 
$600$ & 1.4 \\
NGC\,3003 & 19/02/1999 & CAHA2.2 & CAFOS & CA\#658/10 & $ 1\times1800 $ & 
$800$ & 1.5 \\
 & 19/02/1999 & CAHA2.2 & CAFOS & CA\#658/10 & $ 2\times900 $ & -- & \\
NGC\,3221 & 22/02/1993 & ESO2.2 & EFOSC2 & ESO\#439 & $ 1\times 1200 $ & 
$600$ & \\
 & 22/02/1993 & ESO2.2 & EFOSC2 & ESO\#439 & $ 1\times900 $ & -- & \\
NGC\,3365 & 21/02/1999 & CAHA2.2 & CAFOS & CA\#658/10 & $ 2\times1800 $ & 
$720$ & \\
NGC\,3501 & 22/02/1999 & CAHA2.2 & CAFOS & CA\#658/10 & $ 2\times1800 $ & 
$720$ & \\
NGC\,3600 & 20/02/1999 & CAHA2.2 & CAFOS & CA\#658/10 & $ 2\times1800 $ & 
$720$ & \\
NGC\,3628 & 08/05/1991 & NTT & EMMI & ESO\#596 & $ 1\times1800 $ & $300$ & \\
NGC\,3877 & 20/02/1999 & CAHA2.2 & CAFOS & CA\#658/10 & $ 2\times1800 $ & 
$600$ & \\
NGC\,3936 & 10/02/1995 & NTT & EMMI & ESO\#597 & $ 1\times1200 $ & $300$ & \\
ESO\,379--6 & 09/07/1999 & DAN1.54 & DFOSC & ESO\#697 & $ 1\times3600 $ & 
$2\times600$ & 1.4 \\
NGC\,4206 & 22/02/1999 & CAHA2.2 & CAFOS & CA\#665/17 & $ 2\times1800 $ & 
$720$ & 3.0 \\
NGC\,4216 & 19/02/1999 & CAHA2.2 & CAFOS & CA\#665/17 & $ 2\times1800 $ & 
$600$ & \\
NGC\,4235 & 20/02/1999 & CAHA2.2 & CAFOS & CA\#665/17 & $ 2\times1800 $ & 
$800$ & 1.3 \\
NGC\,4256 & 21/02/1999 & CAHA2.2 & CAFOS & CA\#665/17 & $ 2\times1800 $ & 
$720$ & \\
NGC\,4388 & 20/02/1999 & CAHA2.2 & CAFOS & CA\#665/17 & $ 2\times1200 $ & 
$500$ & \\
NGC\,4700 & 01/08/2000 & DAN1.54 & DFOSC & ESO\#693 & $ 2\times2700 $ & 
$2\times600$ & 1.3 \\
NGC\,4945 & 09/07/1999 & DAN1.54 & DFOSC & ESO\#693 & $ 1\times2700 $ & $600$ 
& 1.6 \\
NGC\,5290 & 21/02/1999 & CAHA2.2 & CAFOS & CA\#665/17 & $ 2\times1800 $ & 
$720$ & \\
NGC\,5297 & 22/02/1999 & CAHA2.2 & CAFOS & CA\#658/10 & $ 1\times1800 $ &
$720$ & 2.0 \\
& 22/02/1999 & CAHA2.2 & CAFOS & CA\#658/10 & $1\times1300$  & -- & \\
NGC\,5775 & 07/05/1991 & NTT & EMMI & ESO\#595 & $ 1\times1800 $ & $600$ & \\
ESO\,274--1 & 01/08/2000 & DAN1.54 & DFOSC & ESO\#693 & $ 2\times2700 $ & 
$2\times600$ & 1.4 \\
NGC\,5965 & 09/08/1999 & CAHA2.2 & CAFOS & CA\#665/17 & $ 2\times2700 $ & 
$120+300$ & \\
NGC\,6722 & 09/07/1999 & DAN1.54 & DFOSC & ESO\#697 & $ 2\times2700 $ & $600$ 
& \\
IC\,4837A & 08/07/1999 & DAN1.54 & DFOSC & ESO\#697 & $ 2\times2700 $ & $600$ 
& 1.9 \\
ESO\,142--19 & 10/07/1999 & DAN1.54 & DFOSC & ESO\#697 & $ 2\times2700 $ & 
$600$ & 1.8 \\
%\noalign{\smallskip}
\hline
\end{tabular}
\end{minipage}
\end{flushleft}
\end{table*}   

\begin{table*}
\setcounter{table}{2}
\caption[continued]{continued}
%\label{T1}
\begin{flushleft}
\begin{minipage}{20cm}\small
\begin{tabular}{lccccccc}
\noalign{\smallskip}
\hline\hline
Galaxy & Date & Telescope & Instrument & H$\alpha$ filter Id. & 
$t_{\rm exp}$\,(H$\alpha$)\,[s] & $t_{\rm exp}$\,(R)\,[s] & Seeing [$''$]\\
\hline
%\noalign{\smallskip}
IC\,4872 & 09/07/1999 & DAN1.54 & DFOSC & ESO\#693 & $ 2\times2700 $ & $600$ 
& 1.7 \\
NGC\,6875A & 10/07/1999 & DAN1.54 & DFOSC & ESO\#697 & $ 2\times2700 $ & 
$600$ & 1.4 \\
MCG-01-53-012 & 10/08/1999 & CAHA2.2 & CAFOS & CA\#665/17 & $ 2\times2700 $ & 
$120+300$ & 1.3\\
IC\,5052 & 08/07/1999 & DAN1.54 & DFOSC & ESO\#693 & $ 2\times2700 $ & $600$ 
& 1.8 \\ 
IC\,5071 & 31/07/2000 & DAN1.54 & DFOSC & ESO\#697 & $ 2\times2700 $ &
$2\times600$ & 1.9 \\
IC\,5096 & 12/11/1999 & DAN1.54 & DFOSC & ESO\#697 & $ 2\times2700 $ & 
$120+300$ & 1.2 \\
NGC\,7064 & 12/11/1999 & DAN1.54 & DFOSC & ESO\#693 & $ 2\times2700 $ & 
$120+600$ & 1.5 \\
NGC\,7090 & 09/07/1999 & DAN1.54 & DFOSC & ESO\#693 & $ 1\times1800 $ & $600$ 
& 1.7 \\
UGC\,11841 & 11/08/1999 & CAHA2.2 & CAFOS & CA\#665/17 & $ 2\times2700 $ & 
$120+600$ & \\
NGC\,7184 & 30/07/2000 & DAN1.54 & DFOSC & ESO\#697 & $ 2\times2700 $ & 
$450+600$ & 1.5 \\
IC\,5171 & 31/07/2000 & DAN1.54 & DFOSC & ESO\#697 & $ 2\times2700 $ & 
$2\times600$ & 1.9 \\ 
IC\,5176 & 10/07/1999 & DAN1.54 & DFOSC & ESO\#693 & $ 2\times2700 $ & $600$ 
& 1.8 \\
NGC\,7339 & 08/08/1999 & CAHA2.2 & CAFOS & CA\#658/10 & $2\times2700$ & 
$120+300$ & 1.7 \\
NGC\,7361 & 31/07/2000 & DAN1.54 & DFOSC & ESO\#693 & $ 2\times2700 $ & $600$ 
& 1.7 \\
NGC\,7412A & 13/11/1999 & DAN1.54 & DFOSC & ESO\#693 & $ 2\times2700 $ & 
$120+600$ & 1.2 \\
UGC\,12281 & 10/08/1999 & CAHA2.2 & CAFOS & CA\#665/17 & $ 2\times2700 $ &
$120+300$ & 1.3 \\
NGC\,7462 & 11/11/1999 & DAN1.54 & DFOSC & ESO\#693 & $ 2\times2700 $ & 
$120+600$ & 1.8 \\
UGC\,12423 & 09/08/1999 & CAHA2.2 & CAFOS & CA\#665/17 & $ 1\times2700 $ & 
$120$ & \\
NGC\,7640 & 10/08/1999 & CAHA2.2 & CAFOS & CA\#658/10 & $ 2\times2700 $ & 
$120+300$ & \\
ESO\,240--11 & 30/07/2000 & DAN1.54 & DFOSC & ESO\#697 & $ 2\times2700 $ & 
$2\times600$ & 2.0 \\
%\noalign{\smallskip}
\hline
\end{tabular}
\end{minipage}
\end{flushleft}
\end{table*}       

As already mentioned our extended survey, which includes the 9 previously 
investigated galaxies presented in an earlier work \citep{Ro00}, consists of 
a grand total of 74 galaxies. 30 galaxies, that is almost 41\% of our survey, 
show extraplanar DIG features (either a pervasive layer, and/or filaments, or 
plumes, etc.). Excluding the 9 galaxies (first sub--sample), there are still 
24 galaxies out of 65 with eDIG detections left, that is $\sim37$\% of the 
survey. As we were primarily aiming to trace the fainter end of the SF 
activity, we were selecting galaxies with a broader range of $L_{\rm FIR}$, 
compared to the galaxies, which were studied earlier by \citet{Le95}, which 
were selected only on the basis of being infrared warm galaxies 
(S$_{60}$/S$_{100}\geq0.4$). Therefore, it is not surprising that we did 
detect a good fraction of galaxies with no extraplanar emission.

\subsection{Results and notes on selected galaxies}

In this subsection a brief description of the observed eDIG morphology 
of selected galaxies is presented in addition to some background information, 
relevant in the DIG context for those targets. The galaxies are 
listed according to their increasing Right Ascension. The H$\alpha$ images 
of all survey galaxies are shown together with the accompanying broad band 
images (R--band, and unsharp--masked R--band image) in Figs.~22--54 
(available only in electronic form at EDP Sciences). However, some 
enlargements of selected galaxies with characteristic and spectacular 
eDIG morphology are included in logarithmic scale as separate figures in 
this section, in order to highlight some finer details of eDIG emission. A 
comparison with observations by other researchers from various wavelength 
regimes in the context of the disk--halo interaction is made, whenever such 
observations were available in the literature. 

\begin{center} {\em \object{NGC\,24}} \end{center} 
The H$\alpha$ morphology of NGC\,24 comprises of planar DIG which is 
visible between several bright \hbox{H\,{\sc ii}} regions in the disk. No 
extraplanar DIG is detected which is basically due to the fact that this 
galaxy is not perfectly edge--on. \citet{Gu92} lists an inclination of 
78$^\circ$. Although the $\rm{L_{FIR}/D^2_{25}}$ ratio (for a definition 
of this expression please cf. Paper~I) is quite low, the $S_{60}/S_{100}$ 
ratio is moderate, and as the H$\alpha$ distribution implies there is 
considerable star formation activity all over the disk. Additional 
evidence comes from a UV--study of nearby galaxies, where the morphology of 
the nuclear region in NGC\,24 was studied \citep{Mao96}. They classify 
NGC\,24 as a galaxy with star-forming morphology, and several knots or 
compact sources can be identified on the UV ($\sim$2300\,{\AA}) image, 
obtained with the Faint Object Camera (FOC) on--board HST. These regions are 
either compact star clusters or individual OB stars.

\begin{center} {\em \object{UGC\,260}} \end{center}
UGC\,260 is shown with another smaller edge--on galaxy, CGCG\,434--012, 
located $2\farcm4$ west of UGC\,260 (see Fig.~21), which has a similar 
redshift ($\Delta v = 47\,\rm{km\,s^{-1}}$). \citet{Re96} included them in 
their list of tidally--triggered disk thickening galaxies. Indeed, the 
morphology in the continuum--subtracted H$\alpha$ image looks quite 
distorted. The \hbox{H\,{\sc ii}} regions are not aligned and just below the 
disk in the very northern part seems to be a small additional galaxy 
possibly in the process of merging, or this represents debris tidal tails. 
Extraplanar DIG is detected, representing a faint layer, with a few 
individual emission patches. The NED lists another galaxy pair between 
UGC\,260 and CGCG\,434--012, which is barely visible in our R--band image. 
We cannot rule out in this particular case, that the eDIG emission is 
triggered by interaction of one of the nearby galaxies.

\begin{center} {\em \object{MCG-2-3-16}} \end{center}
This edge--on galaxy is paired with another edge--on spiral, MCG-2-3-15. 
However, these two galaxies are not physically associated, as MCG-2-3-15 
has a much higher radial velocity of $v_{\rm rad}=5765\,{\rm km\,s^{-1}}$. 
Hence its H$\alpha$ emission is shifted outside the passband of the used 
H$\alpha$ filter. MCG-2-3-16 on the other side is not very prominent 
in H$\alpha$. It is one of our 12 survey galaxies, which has no detectable 
FIR flux at 60$\mu$m or 100$\mu$m at the sensitivity of IRAS. A small arc 
is seen in the western portion, extending about 430\,pc north of the disk, 
which is not extraplanar. The disk shows a slight asymmetry in thickness 
from east to west which might be a projection effect due to a slight 
deviation from the edge--on character. 

\begin{center} {\em \object{UGC\,1281}} \end{center}
UGC\,1281 was included in the Effelsberg/VLA radio continuum survey by 
\citet{Hum91}. However, no radio halo was detected. The H$\alpha$ 
image did not reveal any extraplanar DIG emission. UGC\,1281 has also 
not been detected by IRAS, so this altogether hints for a low star formation 
activity. However, the H$\alpha$ images show several bright 
\hbox{H\,{\sc ii}} regions in the disk which are not aligned. Two bright 
emission knots are slightly offset from the plane to the south.

\begin{center} {\em \object{UGC\,2082}} \end{center}
UGC\,2082 is a northern edge--on spiral galaxy, which is located in the 
direction of the NGC\,1023 group \citep{Tu80}. It has been investigated on 
the basis of a HST survey of large and bright nearby galaxies, studied  with 
the FOC in the UV--regime at $\lambda\sim2300$\,{\AA} \citep{Mao96}. However, 
UGC\,2082 has not been detected, which is an indication that the SF activity 
within this galaxy is very low. That is reflected in our H$\alpha$ images as 
well. We do not detect significant emission except some \hbox{H\,{\sc ii}} 
regions in the disk. The diagnostic FIR ratios are very low, too.

\begin{center} {\em \object{ESO\,362--11}} \end{center}
ESO\,362--11 has a $\rm{L_{FIR}/D^2_{25}}$ ratio of $\sim2.6$, which 
indicates a moderate SF activity. In our H$\alpha$ images a weak layer of 
eDIG is found, but no filaments or plumes are detected. ESO\,362--11 is 
listed in the catalog of Southern Peculiar Galaxies and Associations 
\citep{Ar87} as AM\,0514--370, possibly due to a chain of 4 galaxies which 
are located in the vicinity of ESO\,362--11. These are on the outskirts of 
our covered field, and therefore are not shown in our appendix image. 
\citet{Te88} report that galaxy interactions enhance the efficiency of SF 
activity, which is very likely. However, in the case of ESO\,362--11 it is 
not clear, whether or not the chain of galaxies, which are not in the close 
vicinity of ESO\,362--11, are able to enhance the SF activity. \citet{Coz98} 
even have classified ESO\,362--11 on the basis of their {\em Pico Dos Dias 
Survey} as a starburst galaxy, based on FIR spectral indices.

\begin{center} {\em \object{ESO\,209--9}} \end{center}
This little studied nearby, southern edge--on galaxy has both a moderate 
ratio of $\rm{L_{FIR}/D^2_{25}}$, and $S_{60}/S_{100}$. Its H$\alpha$ 
morphology reveals an extended layer, not as prominent as in NGC\,891 or 
NGC\,3044, but still quite intense. The distribution of the H$\alpha$ 
emission is asymmetric which could be a projection effect, if we were looking 
at the very end of the spiral arm to the south along the line of sight, 
whereas the spiral arm to the north could be winded more tightly. Quite 
extraordinary is an additional emission component, which shows up in the 
whole field around ESO\,209--9. We speculate that this emission, which is 
shown in its fully covered extent in the H$\alpha$+continuum image (see 
Fig.~3), is of Galactic origin. Quite remarkably, there is no hint of any 
filamentary emission in the broad R--band image. Our first impression 
was that this could be straylight from a bright star just outside the 
covered CCD field. However, as there was neither a shift (after alignment) 
nor a change in the emission pattern noticed in the two individual H$\alpha$ 
images, which were offset by $20''$ from one another, and furthermore such 
structures were never recorded prior or after these exposures in other 
object frames, we suspect that these structures indeed have a Galactic 
origin. A visual inspection of both the blue and red DSS images did not 
reveal anything conclusive. Unfortunately, the IR DSS plate of the region 
around ESO\,209--9 is not yet obtained/digitized. The IRAS maps show some 
emission in that area, however, the resolution is not sufficient enough to 
resolve these structures clearly. A possible explanation could be, that this 
emission is Galactic cirrus. This would not be unreasonable, since 
ESO\,209--9 has a galactic latitude of only $b \approx -11^\circ$. 
Unfortunately, this position is just outside the regions studied by the 
AAO/UKST H$\alpha$ 
survey\footnote{http://www.roe.ac.uk/wfau/halpha/halpha.html} of the southern 
galactic plane \citep[e.g.,\,][]{Pa99}, which would have been a good 
test for comparison. It could also be extended red emission (ERE) in the 
diffuse interstellar medium. This ERE was found for high--galactic cirrus 
clouds by \citet{Sz98}. They found a peak of cirrus ERE at 
$\lambda\sim$6000\,{\AA}. However, in order to unravel the true nature of 
this filamentary emission, this should be re--investigated by independent 
deep H$\alpha$ images using a different instrumental setup, and supplemental 
spectroscopy. Fortunately, very recently, we downloaded an available 
H$\alpha$ image from the Southern H$\alpha$ Sky Survey Atlas 
(SHASSA)\footnote{http://amundsen.swarthmore.edu/SHASSA/} (see Fig.~2), which 
became available very recently. For details on SHASSA we refer to 
\citet{Gau01}. Indeed, the field around the galaxy ESO\,209--9 shows 
detectable H$\alpha$ emission, and the H$\alpha$ morphology is clearly 
recovered, despite the spatial resolution of the SHASSA is much lower than in 
our images. 

\begin{figure}[]
\begin{center}
%\rotatebox{0}{\resizebox{8.7cm}{!}{\includegraphics{h4307F2new.ps}}}
\caption[]{\small Field No.~51 of the Southern H$\alpha$ Sky Survey Atlas 
(SHASSA), which includes the Galactic region in the direction of ESO\,209--9. 
The field measures 13$\sq{\degr}$. The field around ESO\,209--9 is indicated 
by a circle.}
\end{center}
\end{figure}

\begin{figure}[]
\begin{center}
%\rotatebox{0}{\resizebox{8.7cm}{!}{\includegraphics{h4307newF3.ps}}}
\caption[]{\small Field around ESO\,209--9, showing very structured, 
cirrus--like galactic emission in this H$\alpha$ image. The corresponding 
R--band image does not show anything filamentary in the field of view. North 
is to the top and East is to the left. The field of view measures  
$\sim13\sq\arcmin$. Orientation: N is to the top and E to the left.}
\end{center}
\end{figure}

\begin{center} {\em \object{NGC\,3003}} \end{center}
NGC\,3003 has been investigated spectroscopically by \citep{Ho95} in 
search of {\em dwarf} Seyfert nuclei, who concluded, based on a detected 
broad emission complex centered at $\lambda$4650{\AA}, that this galaxy is a 
Wolf--Rayet galaxy. NGC\,3003 has a modest $\rm{L_{FIR}/D^2_{25}}$ ratio 
(1.65), however, the $S_{60}/S_{100}$ ratio of $\sim$0.34 hints that there 
is some SF activity due to enhanced dust temperatures. Furthermore, due to 
the slight deviation from its edge--on character, no extraplanar emission 
can be identified reliably. The H$\alpha$ distribution, however, reveals 
strong planar DIG, consistent with the observations by \citet{Hoo99}.
Several bright emission knots can be discerned. About four decades ago a SN 
has been detected $34''$E, and $17''$N of the nucleus (SN\,1961\,F). 
 
\begin{center} {\em \object{NGC\,3221}} \end{center}
This is one of the galaxies with the highest $\rm{L_{FIR}/D^2_{25}}$ ratio 
in our survey (13.8). The $S_{60}/S_{100}$ ratio is also quite high (0.37), 
suggesting enhanced SF activity. Indeed, there are many bright 
\hbox{H\,{\sc ii}} regions visible in the disk, preferentially located in 
the spiral arms, which are slightly recognizable, due to the small deviation 
from its edge--on inclination. A faint eDIG layer is visible. There is a 
slight asymmetry between the northern part and the southern part of the 
galaxy (i.e. the two spiral arms) discernable. NGC\,3221 has also been 
investigated in the radio regime (radio continuum), where \citet{Ir00} 
detected extended disk emission. It is noteworthy to mention that a SN has 
been detected (SN\,1961\,L) in NGC\,3221.  

\begin{center} {\em \object{NGC\,3501}} \end{center}
NGC\,3501 was observed with the Effelsberg 100m radio dish at 5\,GHz in a 
survey to detect radio halos in edge--on galaxies, related to SF driven 
outflows \citep{Hum91}. No extended radio emission has been detected. 
The optical appearance, based on our H$\alpha$ imaging reveals also a 
quiescent galaxy, with no extraplanar emission. Besides some modest 
\hbox{H\,{\sc ii}} regions in the disk, its morphology looks rather 
inconspicuous. In fact, it is also one of the 12 galaxies in our survey 
which has no FIR detections.  

\begin{figure}[h]
\begin{center}
%\rotatebox{0}{\resizebox{8.7cm}{!}{\includegraphics{h4307newF4.ps}}}
\caption[]{\small Extraplanar DIG layer in NGC\,3600. The eDIG morphology 
is reminiscent of a starburst galaxy. The field of view measures 
$\sim$11.3\,kpc$\times$11.3\,kpc. Orientation: N is to the top and E to the 
left.}
\end{center}
\end{figure}

\begin{center} {\em \object{NGC\,3600}} \end{center}
This edge--on spiral galaxy has been included in the catalog of Markarian 
galaxies as Mrk\,1443 \citep{Maz86}. The broadband R image shows a prominent 
bulge and indicates a slight warp which was also noticed by \citep{Sa90}. In 
our H$\alpha$ image it becomes even more obvious that the disk is warped. 
There is some extraplanar DIG detected, basically around the central 
regions, where extended emission seems to be ejected from the nuclear 
region. There is, besides the nucleus, one bright emission region, 
presumably consisting of several smaller components, and several fainter 
knots visible in the disk. Although the $\rm{L_{FIR}/D^2_{25}}$ is quite 
moderate (1.4), NGC\,3600 has a relatively high $S_{60}/S_{100}$ ratio of 
$\sim0.44$. Spectroscopy of the nuclear region reveals strong emission from 
the Balmer lines (H$\alpha$, H$\beta$), as well as from several forbidden 
low ionization lines, such as [\hbox{N\,{\sc ii}}], and [\hbox{S\,{\sc ii}}] 
\citep{Ho95}. We show a detailed view of NGC\,3600 in Fig.~4.

\begin{center} {\em \object{NGC\,3628}} \end{center}
NGC\,3628 is a member of the {\em Leo triplet}, which also includes 
NGC\,3623 and NGC\,3627. It is a starburst galaxy with a very prominent dust 
lane, which obscures most parts of the emission in the galactic midplane. It 
has been studied extensively in almost all wavelength regimes, and has also 
been the target for a multi--wavelength study in the context of the 
disk--halo connection. Radio continuum observations \citep{Sc84} reveal 
extended emission, and in the X--ray regime a T$\sim2\times10^6$\,K, 
extended halo has been detected by sensitive PSPC observations with ROSAT 
\citep{Da96}. Prominent X--ray emission, tracing the collimated outflow 
from the nuclear starburst, was found to be spatially correlated with the 
H$\alpha$ plume \citep{Fa90}. Extraplanar dust has already been detected 
\citep{How99,Ro01}. We also observed extended emission from the nuclear 
outflow, localized extended emission (filamentary), and several extraplanar 
\hbox{H\,{\sc ii}} regions.

\begin{center} {\em \object{NGC\,3877}} \end{center}
NGC\,3877 is a northern edge--on spiral galaxy, which ranks among the top 15 
galaxies of our survey, according to the SFR per unit area. A SFR from 
observations in the UV has been derived, which yielded $0.55\,{\rm 
M_{\odot}\,yr^{-1}}$ \citep{Don87}. A nuclear spectrum reveals quite 
strong emission of H$\beta$ \citep{Ho95}, whereas H$\alpha$, and the 
[\hbox{N\,{\sc ii}}] lines are also clearly detected. Just recently a SN of 
type II has been detected in NGC\,3877, namely SN\,1998\,S \citep{Fi98}. In 
Fig.~6 we show our broadband R image, which was obtained almost a year after 
its discovery. The SN (marked by a circle), although considerably fainter, 
is still visible in our image. \citet{Ni95} report a $\lambda2.8$\,cm flux 
for NGC\,3877 of $S_{\rm tot}=23\pm8$\,mJy. However, no radio map was shown, 
and they did not comment further on this galaxy. The H$\alpha$ morphology, 
as seen in our images, reveal extended emission with some small filaments 
and plumes. Quite remarkable is the distribution of very strong 
\hbox{H\,{\sc ii}} regions and knots which are seen all over the disk 
(cf.~Fig.~5). The nuclear region is very compact in H$\alpha$.

\begin{figure}[]
\begin{center}
%\rotatebox{0}{\resizebox{8.7cm}{!}{\includegraphics{h4307newF5.ps}}}
\caption[]{\small Extraplanar DIG layer in NGC\,3877 seen in this continuum 
subtracted H$\alpha$ image. There is a pervasive eDIG layer clearly visible.
The field of view measures 12\,kpc$\times$12\,kpc. Orientation: N is to the 
top and E to the left.}
\end{center}
\end{figure}

\begin{figure}[]
\begin{center}
%\rotatebox{0}{\resizebox{8.7cm}{!}{\includegraphics{h4307newF6.ps}}}
\caption[]{\small SN\,1998\,S (marked by a circle) in NGC\,3877. Our R--band 
image was obtained 354 days after the discovery of SN\,1998\,S. Image size 
and orientation is same as above.}
\end{center}
\end{figure}

\begin{center} {\em \object{NGC\,4216}} \end{center}
Another Virgo cluster spiral, NGC\,4216 has been classified as a WR--galaxy, 
based on detected \hbox{He\,{\sc ii}}\,$\lambda$4686 emission from a few 
regions within NGC\,4216 \citep{Sc99}. \citet{Po89b} investigated 
the morphology of the nuclear emission, and concluded it as diffuse, while 
the circumnuclear region was characterized as faint and patchy. There is 
nuclear emission detected from X--rays \citep{Fa92}. \citet{Mao96} did not 
detect UV emission in NGC\,4216 in their HST survey. Our H$\alpha$ image did 
not reveal extraplanar emission. The nucleus is the strongest source, whereas 
several smaller \hbox{H\,{\sc ii}} regions in the outer spiral arm 
contribute also to the H$\alpha$ emission.

\begin{center} {\em \object{NGC\,4235}} \end{center}
NGC\,4235 has been classified as a Seyfert galaxy of type 1 \citep{We78},  
and has no physical companion \citep{Dah84}. \citet{Po89a} has studied the 
nuclear environment using narrowband imaging of H$\alpha$ and 
[\hbox{O\,{\sc iii}}]. He found a bright nucleus with an extended region 
towards the NE direction at P.A. $\sim48^\circ$, which extends $\sim4.4''$. 
Neither disk \hbox{H\,{\sc ii}} regions were detected, nor ionized gas above 
the plane. Radio continuum observations \citep{Hum91} also did not find 
evidence for extended emission. In the X--ray regime the strong nuclear 
region is detected with a luminosity of $1.55\times10^{42}\,{\rm erg\,
s^{-1}}$ \citep{Fa92}. In a study of large--scale outflows in edge--on 
Seyfert galaxies \citet{Co96b} found no double peaked line profiles, or any 
evidence for extended line regions, and no minor axis emission, too. However, 
on radio continuum images obtained with the VLA at 4.9\,GHz there was a 
diffuse, bubble--like, extended structure ($\sim9$\,kpc) found in addition 
to the unresolved nucleus \citep{Co96a}. Our H$\alpha$ image reveals a bright 
nucleus with a faint extended layer, which is restricted to the circumnuclear 
part. NE of the nucleus a depression is visible, possibly due to absorbing 
dust, as already noted by \citet{Po89a}, and easily visible in the broad band 
HST image by \citet{Mal98}. This is one of the very few Seyfert galaxies that 
appear in our survey. The role of minor axis outflows in Seyfert galaxies, 
and thus the contribution to the IGM enrichment and heating still has to be 
explored.    

\begin{center} {\em \object{NGC\,4388}} \end{center}
NGC\,4388 has been identified as the first Seyfert\,2 galaxy in the Virgo 
cluster \citep{Ph82}, and has been studied extensively in various 
wavelength regimes, including optical line imaging and spectroscopy 
\citep[e.g.,\,][]{Kee83,Po88,Cor88}, radio continuum 
\citep[e.g.,\,][]{St88,Hu91}, and in the high energy waveband extended soft 
X--ray emission out to a radius of 4.5\,kpc has been reported \citep{Mat94}. 
From the optical morphology and kinematics it was derived that NGC\,4388 
possesses complex gas kinematics and is composed of several nucleated 
emission line regions. A prominent feature reaches out to $\sim18''$ at 
P.A.\,$\sim10^\circ$ \citep{He83}. The radio continuum maps revealed a 
double peaked radio source close to the optical nucleus plus a cloud of 
radio emitting material, apparently ejected from the nucleus 
\citep[e.g.,\,][]{St88,Ir00}. 

Some speculation on the true membership to the Virgo cluster exists, although 
NGC\,4388 is located near the core of the Virgo cluster. This is because of 
its relatively high systemic velocity. However, most authors assume it is a 
member of the Virgo cluster. Therefore, ram pressure stripping seems to play 
a role as NGC\,4388 interacts with the ambient intracluster medium (ICM).
Optical narrowband imaging in H$\alpha$ and [\hbox{O\,{\sc iii}}] has 
revealed that the morphology of the extended ionized gas is composed of two 
opposed radiation cones \citep{Po88}, which give rise to a {\em hidden 
Seyfert\,1} nucleus, as favored in the {\em unification scheme} of AGN. 
Recent investigations of NGC\,4388 using Fabry--Perot imaging techniques 
have revealed a complex of highly ionized gas $\sim4$\,kpc above the disk 
\citep{Ve99}. They found blueshifted velocities of $50-250\,{\rm km\,
s^{-1}}$ NE of the nucleus. Furthermore they assume the velocity of 
the extraplanar gas to be unaffected by the inferred supersonic motion of 
NGC\,4388 through the ICM of the Virgo cluster, and suggest that the galaxy 
and high--$|z|$ gas lies behind the Mach cone.   

Our narrowband imaging of NGC\,4388 (see Fig.~7) reveals also extended 
emission which is pointing away from the galactic disk to the halo in the NE 
direction. Furthermore a faint eDIG layer is visible. Very recently deep 
H$\alpha$ images obtained with the Subaru telescope revealed very 
extended emission--line region in H$\alpha$ and [\hbox{O\,\sc{iii}}] at 
distances of up to $\sim$35\,kpc \citep{Yo02}. 

\begin{figure}[]
\begin{center}
%\rotatebox{0}{\resizebox{8.7cm}{!}{\includegraphics{h4307newF7.ps}}}
\caption[]{\small Continuum subtracted H$\alpha$ image of NGC\,4388, 
revealing an eDIG layer and an extended plume. The displayed area equals 
$\sim$36\,kpc$\times$32\,kpc at the distance of NGC\,4388. Orientation: N 
is to the top and E to the left.}
\end{center}
\end{figure}

\begin{center} {\em \object{NGC\,4700}} \end{center} 
NGC\,4700 has been listed as a \hbox{H\,{\sc ii}} region galaxy in the list 
of \citet{Rod87}, while \citet{Hew91} list it as a Seyfert\,2 galaxy. It 
has been included in the HST imaging survey of nearby AGN \citep{Mal98}. 
NGC\,4700 has a relatively large ratio of $S_{60}/S_{100}$ of $\sim 0.51$, 
making it a promising candidate with a potential DIG halo according to the 
diagnostic DIG diagram. Indeed, a relatively bright, and extended gaseous 
halo with a maximum extent of 2.0\,kpc above/below the galactic plane is 
discovered. One of the filaments can even be traced farther out to 
$\sim3$\,kpc. The morphology of DIG in NGC\,4700 is asymmetrical, with the 
eastern and middle part being most prominent. Five distinct bright 
\hbox{H\,{\sc ii}} region complexes, which are composed of several individual 
smaller regions across the disk, can be discerned. Some prominent filaments 
above active regions protrude from the disk into the halo. A small number of 
bright \hbox{H\,{\sc ii}} regions is seen in this part of the disk, and the 
maximum extent of the eDIG is visible above those regions. The western part 
of the galaxy is lacking in bright \hbox{H\,{\sc ii}} regions, hence the 
suppressed extend of the halo in this western part. NGC\,4700 also bears an 
extended radio halo (radio thick disk), with the maximum extent 
correspondingly on the same position above the disk \citep{Da01}, although 
slightly more extended in the radio continuum image as in our H$\alpha$ 
image (see Figs.~8+9).

\begin{figure}[]
\begin{center}
%\rotatebox{0}{\resizebox{8.7cm}{!}{\includegraphics{h4307F8.ps}}}
\caption[]{\small A detailed H$\alpha$ view of NGC\,4700 showing a bright 
and extended gaseous halo, superposed by individual filaments.}
\end{center}
\end{figure}

\begin{figure}[]
\begin{center}
%\rotatebox{0}{\resizebox{8.7cm}{!}{\includegraphics{h4307F9.ps}}}
\caption[]{\small The extended radio halo of NGC\,4700 at $\nu$=1.4\,GHz 
(depicted in contours), overlaid onto the DSS image \citep{Da01}, courtesy 
M.~Dahlem.}
\end{center}
\end{figure}

\begin{center} {\em \object{NGC\,4945}} \end{center} 
NGC\,4945 is a well studied southern edge--on spiral galaxy, which belongs 
to the {\em Centaurus group}. It has been studied at various wavelengths, 
including the visual, IR, and radio regime. \citet{He90} found evidence for 
a starburst driven superwind in NGC\,4945, and \citet{Mo94} derive a 
$\sim400$\,pc size starburst in addition to the presence of a visually 
absorbed Seyfert nucleus. They conclude that NGC\,4945 is in an advanced 
stage of evolution from a starburst to a Seyfert galaxy. Extended emission 
from the disk was detected by radio continuum observations 
\citep{Ha89,Co96a}, reaching a diameter of $\sim23$\,kpc. X--ray emission 
was detected from the nuclear region, however, no extended emission was found 
in ROSAT PSPC observations \citep{Co98}, probably due to quite large 
absorbing columns (${\rm N_H = 1.5\times10^{21}\,cm^{-2}}$), as NGC\,4945 is 
situated near the galactic plane ($b\sim13^\circ$). This decreases the 
sensitivity for the soft X--ray regime.

NGC\,4945 is one of the few starburst galaxies that we have included in our 
H$\alpha$ survey, and it is no surprise that it has the second highest 
$\rm{L_{FIR}/D^2_{25}}$ ratio of $\sim 14.8$ of our studied galaxies. Already 
investigated by \citet{Le95}, NGC\,4945 shows strong extraplanar DIG with 
many filaments protruding from the disk into the halo. There are at least 
two bright filaments on either side of the disk visible on our narrowband 
images. Several prominent dust patches, which obscure some parts of the 
emission south of the galactic plane is a further characteristic pattern for 
this galaxy. In Fig.~10 we show an enlargement of the middle part, which 
nicely shows the outflow cone from the nuclear starburst. 

\begin{figure}[]
\begin{center}
%\rotatebox{0}{\resizebox{8.7cm}{!}{\includegraphics{h4307newF10.ps}}}
\caption[]{\small A zoom onto the outflowing cone in this H$\alpha$ image 
of NGC\,4945. The displayed portion of this galaxy measures 
$\sim$5.6\,kpc$\times$5.6\,kpc. Orientation: N is to the top and E to the 
left.}
\end{center}
\end{figure}

\begin{center} {\em \object{NGC\,5290}} \end{center} 
This northern edge--on galaxy is part of the LGG\,361 group \citep{Ga93}, 
which teams up with NGC\,5289, which is located $\sim13'$ to the south, and 
has a velocity difference to NGC\,5290 of $\sim67\,\rm{km\,s^{-1}}$ 
\citep{Hu83}. In the optical NGC\,5290's most striking character is a 
box--shaped bulge \citep{Sou87}. The $\rm{L_{FIR}/D^2_{25}}$ ratio is 
moderate ($\sim2.6$), and NGC\,5290 shows some extended emission, where a 
faint layer is detected, and some filaments, basically coming from the 
nuclear region and reaching into the halo, can be discerned (see Fig.~11). 
The morphology that is visible on the H$\alpha$ image resembles those of a 
starburst galaxy, although NGC\,5290 has no starburst--like FIR parameters, 
as the $S_{60}/S_{100}$ ratio is considerably lower (0.31).

\begin{figure}[]
\begin{center}
%\rotatebox{0}{\resizebox{8.7cm}{!}{\includegraphics{h4307newF11.ps}}}
\caption[]{Continuum subtracted H$\alpha$ image of NGC\,5290, showing a 
pervasive eDIG layer and its morphology resembles those of a nuclear 
starburst galaxy. The displayed area measures 
$\sim$28.6,kpc$\times$28.6\,kpc at the distance of NGC\,5290. Orientation: 
N is to the top and E to the left.}
\end{center}
\end{figure}

\begin{center} {\em \object{NGC\,5297}} \end{center} 
Another northern edge--on spiral, NGC\,5297 forms a binary galaxy with 
NGC\,5296 \citep{Tu76}, which is separated by about $1.5'$ from NGC\,5297, 
and has a velocity difference of $\Delta v \approx 360\rm{km\,s^{-1}}$. 
Another interesting source is also located in the direct vicinity, a quasar 
of V=19.3\,mag \citep{Ar76}. This quasar ([HB89]\,1342+440), which has a 
redshift of $z=0.963$, is located $2.5'$ to the SW. Arp reports on a luminous 
extension from NGC\,5236 pointing at the QSO, which \citet{Sh90} did not 
confirm. However, \citet{Sh90} did comment on the unusually bright 
off--center secondary nucleus of NGC\,5296. The QSO is marked by a circle in 
our R--band image in Fig.~40. The outer spiral arms of NGC\,5297 show 
evidence of perturbation by the S0 companion, as reported by \citep{Ra95}. 
   
Radio continuum observations of NGC\,5297 have been performed 
\citep{Hu85,Ir99,Ir00}. While \citet{Ir99} claims extended radio continuum 
emission from NGC\,5297, still, higher resolution observations show only 
very weak emission \citep{Ir00}. Our H$\alpha$ image does not reveal any 
extraplanar emission. Thus, we do not see enhanced SF activity due to the 
interaction with the companion galaxy in this case. The $\rm{L_{FIR}/
D^2_{25}}$ ratio is moderate. We also note, that this galaxy actually is not 
perfectly edge--on. 

\begin{center} {\em \object{NGC\,5775}} \end{center} 
From \hbox{H\,{\sc i}} observations it was found that NGC\,5775 is an 
interacting galaxy with its neighbor face--on spiral NGC\,5774. Emission 
along a bridge has been detected, although no tidal arms are discovered 
\citep{Ir94}. NGC\,5775 is a good studied northern edge--on galaxy in the 
disk--halo context, which has been imaged in H$\alpha$ 
\citep{Le95,Co00,Tu00}, where extraplanar emission was detected out to 
over 5\,kpc above the galactic plane. It is the galaxy with the highest 
$\rm{L_{FIR}/D^2_{25}}$ ratio in our sample and is a starburst--type galaxy, 
although no clear indications exist on a nuclear starburst. A bright halo 
with individual filaments superposed in a prominent {\em X--shape} are 
discovered on our narrowband image, consistent with earlier observations, 
mentioned above. Extraplanar DIG has been detected spectroscopically out to 
9\,kpc \citep{Ra00,Tu00}, and extended radio continuum emission has been 
detected as well \citep{Hum91,Du98}. Extended X--ray emission coexistent 
with a radio continuum spur and optical filament was discovered on ROSAT 
PSPC archival images \citep{Ro01}.
 
\begin{center} {\em \object{ESO\,274--1}} \end{center} 
ESO\,274--1 belongs to the {\em Cen A} group of galaxies, and \citet{Ban99} 
derive a \hbox{H\,{\sc i}} mass of $430\times10^6\,\rm{M_{\odot}}$, ranking 
on position five of the {\em Cen A} group. This little studied southern 
edge--on galaxy was included in the recent list of dwarf galaxies by 
\citet{Cot97}, who studied new discovered dwarf galaxies in the nearest 
groups of galaxies. From optical inspection of the DSS images, it is obvious 
that this galaxy has a low surface brightness. It is the galaxy with the 
lowest $\rm{L_{FIR}/D^2_{25}}$ ratio among the IRAS detected galaxies in our 
sample (0.19)! Quite remarkably however, the $S_{60}/S_{100}$ ratio is among 
the highest in our sample, making ESO\,274--1 a promising candidate in search 
for eDIG. We show in Fig.~12 an enlargement of the most active SF regions, 
which show strong local extended emission. The SW and middle part is most 
prominent, whereas there are regions in the NE part, where this galaxy 
(superposed by a crowded field of foreground stars) is almost invisible in 
H$\alpha$. 

\begin{figure}[t]
\begin{center}
\rotatebox{0}{\resizebox{8.7cm}{!}{\includegraphics{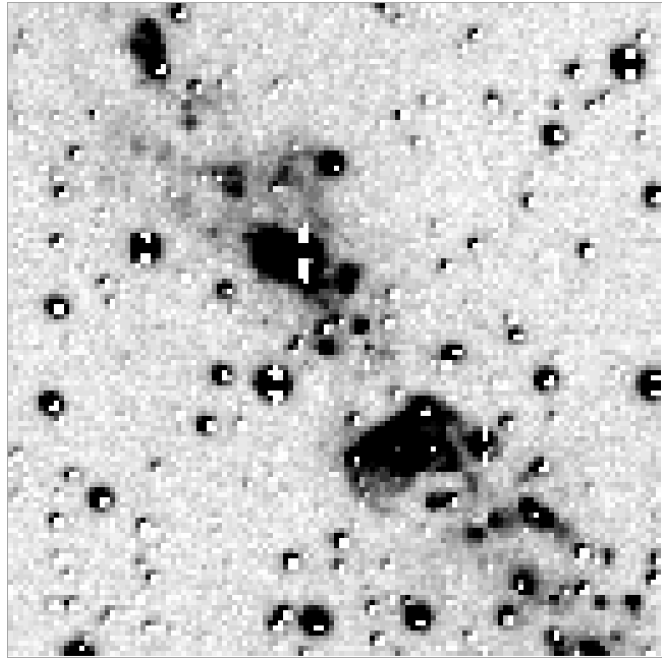}}}
\caption[]{\small Prominent extraplanar emission regions in this continuum 
subtracted H$\alpha$ image of ESO\,274--1. The displayed area equals  
$\sim$4.4\,kpc$\times$4.4\,kpc at the distance of ESO\,274--1. Orientation: 
N is to the top and E to the left.}
\end{center}
\end{figure}

\begin{center} {\em \object{ESO\,142--19}} \end{center} 
This is a perfect edge--on galaxy, and it is the only one of type S0--a in 
our sample. It was only included because of its classification as type Sa in 
the NED. It shows a strong dust lane, and a strikingly bright, and 
extended bulge in the R--band image. An upper radio continuum flux limit of 
20\,mJy at $\lambda$11\,cm has been given by \citet{Sa84}, and a upper flux 
limit of $\rm{F_{[\scriptsize \hbox{N\,{\sc ii}}]} < 1.6\times10^{-14}\,
erg\,s^{-1}\,cm^{-2}}$ has been derived by \citet{Ph86}. We detect 
no extended emission in ESO\,142--19. Even the disk is almost invisible on 
our H$\alpha$ image. 

\begin{center} {\em \object{IC\,5052}} \end{center} 
IC\,5052 is a late--type spiral with classifications Scd or Sd. Radio 
emission detected by $\lambda35$\,cm observations revealed an unusually 
double peaked radio emission complex associated with the disk, which was not 
coming from the nucleus \citep{Ha91}. 

Our H$\alpha$ image reveals a bright layer of extraplanar DIG superposed by 
individual filaments and shells (see Fig.~13). There is a depression in the 
H$\alpha$ distribution noticed in the southern part of the galaxy, due to 
significant dust absorption, which almost separates the galaxy in two parts. 
An asymmetrical distribution is also visible in the R--band image. The 
southern edge is dominated by H$\alpha$ emission, whereas the broadband 
image shows almost no intensity. Several arcs and shells make this galaxy a 
good target for further high resolution studies. 

Pa$\alpha$ observations with NICMOS 2 HST \citep{Boe99} showed excess 
emission of an isolated region in the disk--halo interface, slightly offset 
from the disk, which is possibly associated with one of the brighter 
emission regions in the southern part of our H$\alpha$ image. The diagnostic 
ratios are both moderate, so it is quite remarkable that extended DIG 
emission is detected in that abundance in IC\,5052.  

\begin{figure}[]
\begin{center}
\rotatebox{0}{\resizebox{8.7cm}{!}{\includegraphics{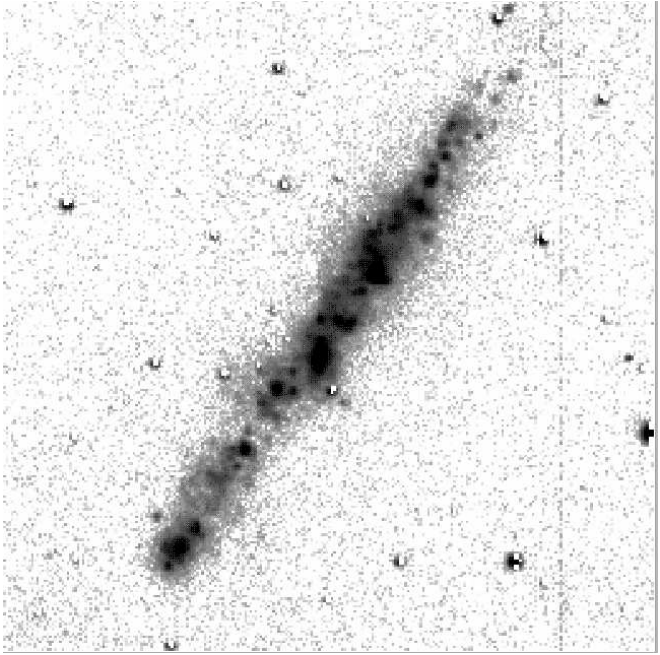}}}
\caption[]{Continuum subtracted H$\alpha$ image of IC\,5052, showing an eDIG 
layer, filaments and shells. The displayed area is 
$\sim$11.8\,kpc$\times$11.8\,kpc. Orientation: N is to the top and E to the 
left.}
\end{center}
\end{figure}

\begin{center} {\em \object{NGC\,7064}} \end{center} 
NGC\,7064 is a \hbox{H\,{\sc ii}} region like galaxy \citep{Ki90}. Radio 
emission has been detected at $\lambda$35\,cm \citep{Ha91}, which extends 
over the inner disk with a maximum intensity at the nucleus. The disk 
emission is more pronounced to the east, which corresponds to the optical 
emission distribution. They further show in their map a weak extension to 
the south, which apparently has no optical counterpart. NGC\,7064 has also 
been detected in the ROSAT All Sky Survey (RASS), which was included in a 
study of IRAS galaxies \citep{Bo98}. The X--ray intensity maximum is 
offset from the IRAS position (nucleus). Furthermore, weaker emission is 
detected to the south of NGC\,7064 at larger distances far out in the halo 
region. Presently it is not clear if it is related to NGC\,7064. In Fig.~14 
we show our H$\alpha$ image, scaled logarithmically, to show the faint halo. 
A few emission patches and plumes are visible to the south of the disk, and 
also in the northern part. The disk emission is asymmetrically distributed, 
with the eastern part being most prominent. There is a prominent hole in the 
southern disk, bisecting the disk almost entirely. In our logarithmically 
scaled image the hole is almost filled with diffuse emission, but its 
intensity is weaker than in the eastern and western part of the disk--halo 
interface. Although the $\rm{L_{FIR}/D^2_{25}}$ ratio is quite low, it 
should be noted that the $S_{60}/S_{100}$ ratio is the highest in our sample.

\begin{figure}[]
\begin{center}
\rotatebox{0}{\resizebox{8.7cm}{!}{\includegraphics{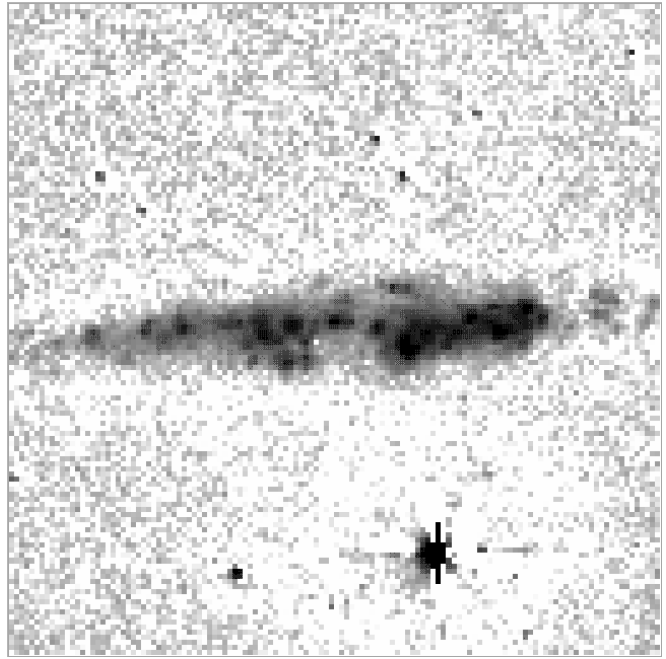}}}
\caption[]{Continuum subtracted H$\alpha$ image of NGC\,7064, revealing 
a faint extraplanar DIG layer. The displayed area measures 
$\sim$10.0\,kpc$\times$10.0\,kpc. Orientation: N is to the top and E to 
the left.}
\end{center}
\end{figure}

\begin{center} {\em \object{NGC\,7090}} \end{center} 
This southern edge--on spiral has radio continuum detections at $\lambda$
35\,cm \citep{Ha85}. The radio emission follows the plane of the galaxy, 
and extends above it in two distinct spurs out to a distance of 
$\sim1.5$\,kpc. The radio peak coincide with the optical nucleus, however, 
they note no obvious nuclear radio source. Our R--band image shows that a 
prominent dust lane with patchy structures runs across most parts of 
the disk offset slightly to the north. It is very irregular, unlike most 
other prominent dust lanes, such as in NGC\,891 or IC\,2531. The H$\alpha$ 
image (Fig.~15) reveals a faint halo, in addition to some filaments and 
extended emission in the form of knots. The northeastern part of the galaxy 
has the highest intensity in emission, whereas the southern half is almost 
absent, except very few emission patches (\hbox{H\,{\sc ii}} regions) in the 
disk. The filaments protrude basically from the nuclear region into the 
halo. It should be noted that the extended radio continuum emission is 
coincident with the H$\alpha$ extended emission.

\begin{figure}[]
\begin{center}
\rotatebox{0}{\resizebox{8.7cm}{!}{\includegraphics{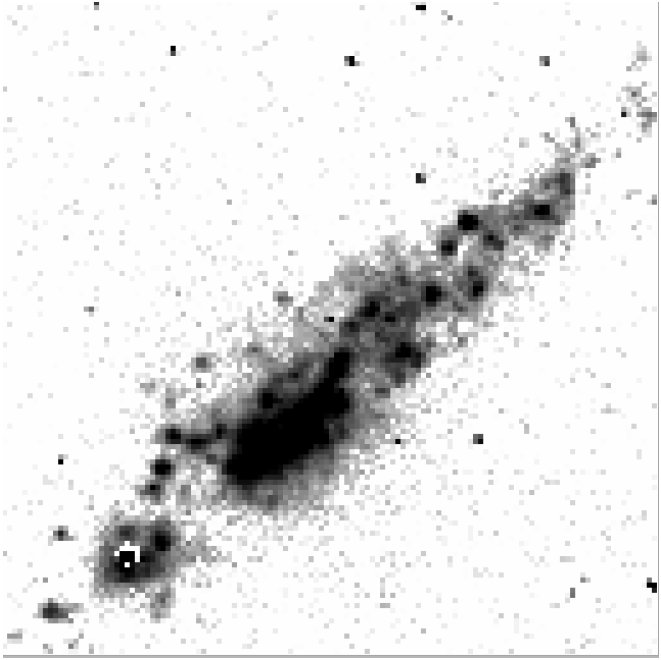}}}
\caption[]{Continuum subtracted H$\alpha$ image of NGC\,7090, showing an 
eDIG layer, with filaments and plumes. The displayed area of the galaxy 
measures $\sim$7.8\,kpc$\times$7.8\,kpc. Orientation: N is to the top and E 
to the left.}
\end{center}
\end{figure}

\begin{center} {\em \object{NGC\,7184}} \end{center} 
NGC\,7184 was detected in the radio regime at $\lambda$20\,cm continuum 
emission \citep{Con87}, and bears a double peaked brightness distribution. 
This was confirmed by \citet{Ha91}, who detected two maxima at 
$\lambda$35\,cm. NGC\,7184 is located at a projected distance of 163\,kpc 
($20.9'$) to NGC\,7185, as quoted by \citet{Ko89}, who studied the group 
environment of Seyfert galaxies. A peculiar supernova was detected in 
NGC\,7184, named SN\,1984\,N \citep{Bar99}. Our H$\alpha$ image did not 
show any extraplanar emission, which might be due to the fact, that 
NGC\,7184 is not perfectly edge--on. The outstanding feature in NGC\,7184 is 
the inner ring, which is also prominently seen on the R--band image. There 
is a well pronounced sub--structure visible, and several \hbox{H\,{\sc ii}} 
regions can be discerned in the ring and in the outer spiral arms. 

\begin{center} {\em \object{NGC\,7339}} \end{center} 
The late type spiral NGC\,7339 is accompanied by the galaxy NGC\,7332 at a 
projected distance of $5.2'$, the latter one being a peculiar S0 galaxy, 
which has a velocity difference of $\Delta v = -141\,\rm{km\,s^{-1}}$ in 
comparison to NGC\,7339. A supernova (SN\,1989\,L) has been discovered in 
NGC\,7339 \citep{Bar99}. Several bright, and compact knots are 
seen in our H$\alpha$ image, but no extraplanar emission is detected. 
NGC\,7332, which is also visible on our frame, is completely absent in 
H$\alpha$.

\begin{center} {\em \object{UGC\,12281}} \end{center} 
UGC\,12281 has already been studied in the DIG context a few years ago by 
\citet{Pi94}. They claim to have detected two extraplanar emission 
line features (discrete clouds), and a clumpy plume. We can confirm the 
presence of these two clouds, and the plume is also marginally detected in 
our image, which suffers from low S/N. 

\begin{center} {\em \object{NGC\,7462}} \end{center} 
This southern edge--on spiral was detected with the VLA at $\nu$=1.49\,GHz 
\citep{Cons87}, and \citet{Mao96} have conducted UV observations at 
($\lambda2300$\,{\AA}) of the inner $22''\times22''$ region with the HST, 
where they detected a {\em star-forming} morphology in several distinct 
regions (knots). Our H$\alpha$ image (Fig.~16) reveals an extended DIG layer 
with individual plumes and faint filaments superposed. The disk emission is 
composed of several distinct emission complexes, clustered basically in three 
regions (nucleus + two regions within the disk), and some fainter regions in 
the outskirts of the disk.

\begin{figure}[]
\begin{center}
\rotatebox{0}{\resizebox{8.7cm}{!}{\includegraphics{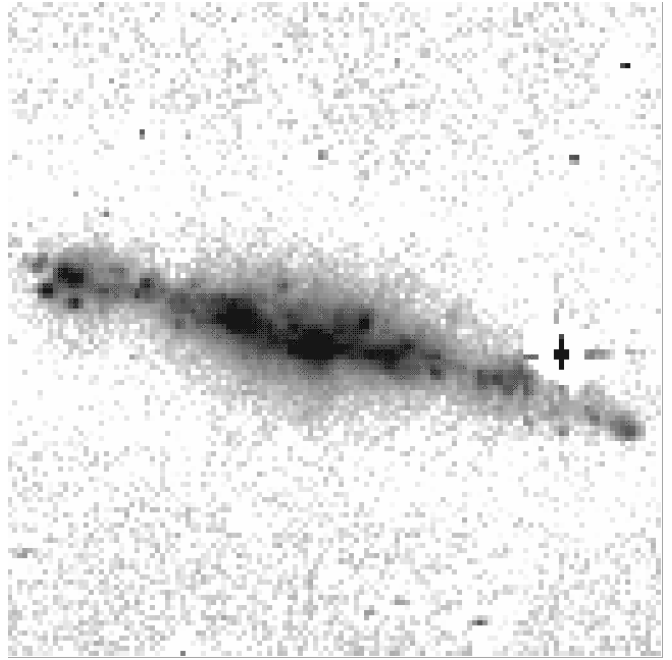}}}
\caption[]{Continuum subtracted H$\alpha$ image of NGC\,7462, showing 
extraplanar emission. The displayed area equals $\sim$10\,kpc$\times$10\,kpc 
at the distance of NGC\,7462. Orientation: N is to the top and E to the left.}
\end{center}
\end{figure}

\begin{center} {\em \object{UGC\,12423}} \end{center} 
UGC\,12423 belongs to the {Pegasus I} cluster of galaxies, and was once 
reported being one of the most massive and luminous spirals, with a derived 
\hbox{H\,{\sc i}} mass of $\sim9\times10^{9}\,\rm{M}_{\odot}$, and $\log 
\rm{M_H/L_B} = +0.14$ \citep{Bo82,Sc83}. \citet{Hum91} did not detect 
extended emission from UGC\,12423 in their radio continuum survey. 
Our H$\alpha$ image also does not show any extraplanar emission. The galaxy 
is almost invisible in H$\alpha$. This can partly be attributed to 
insufficient S/N, as we unfortunately only acquired one exposure for 
H$\alpha$, and also the R--band image was not long enough integrated. This 
needs to be re--investigated with more sensitive observations. 

\subsection{H$\alpha$ sensitivities and error budgets of the continuum 
subtraction}

From the direct comparison of the distribution of eDIG in the galaxies in 
our survey, with those already observed by \citet{Le95} and \citet{Ra96}, we 
can conclude that we have reached similar and in a few cases somewhat better 
sensitivities. This is mostly true for the galaxies, which were obtained 
with DFOSC at La Silla. The Calar Alto observations, however, have been 
acquired through relatively wide H$\alpha$ filters in several cases 
($\Delta\lambda\sim168$\,{\AA}), so the sensitivity is not as good as for 
the La Silla observations, which were made through H$\alpha$ filters of 
$\Delta\lambda\sim62$\,{\AA}. The estimated mean sensitivities of the 
galaxies observed with DFOSC are of the order of a few $\rm{cm^{-6}\,pc}$. 

Typically there are some uncertainties associated with the scaling procedure, 
to be applied for the continuum subtraction. The scaling factor has been 
determined from the ratio of the H$\alpha$ and continuum countrates of 
individual stars in the object frames. Of course, there arise uncertainties 
as the actual continuum spectrum of a galaxy will be different from that of 
individual stars, as the galactic continuum is the superposition of all 
stars of the underlying stellar population. However, practice has shown, 
that this is a good method and the uncertainties typically are of the order 
of 5--10\,\%. A careful analysis, however, is necessary in order to match 
the ideal scaling factor. The determined factor from the intensity ratio of 
the stars in the H$\alpha$ and continuum frame has been slightly changed in 
a few cases where the derived scaling factor did not seem to match perfectly, 
as artifacts were observed in the bulge region, which is a good indicator 
for the accuracy of the scaling process. Then the galaxy profile was 
analyzed carefully, as the scaling factor was changed in very small steps to 
find the optimal (in a somewhat conservative manner) value, not to under- 
and oversubtract the emission. This is a well established procedure 
\citep[for more details see e.g.,\,][]{Hoo99,Ro00}. 

Alternatively, it is possible to determine the scaling factor using 
continuum regions in the galaxy. This is something which would work much 
better for {\it face--on} galaxies, but is somewhat more difficult for 
{\it edge--on galaxies}, as here occasionally the prominent dust lane makes 
it difficult to assess whether there is underlying emission, which is strong 
enough to be seen after subtracting the continuum. Furthermore, the true 
H$\alpha$ emission is only revealed after the continuum subtraction process. 
Hence, knowing regions free of emission is somewhat difficult to judge, 
prior to the subtraction process. However, testing this method, applied to a 
few individual galaxies, have shown that there is a good agreement between 
using stars or continuum light within the galaxies for the determination of 
the scaling factor. 

\subsection{Discussion}

This first large H$\alpha$ survey of edge--on galaxies has shown that eDIG is 
a general phenomenon {\em only} for galaxies which exceed a minimal threshold 
of SF activity. As already suggested by \citet{Ra96} and \citet{Ro00}, the 
results presented in this work corroborate this view that a considerable 
fraction of late--type spirals indeed show extended DIG emission in their 
galactic halos. Although not as common as in starburst galaxies eDIG is a 
widespread phenomenon and is not a characteristic belonging only to a small 
fraction of late--type spirals. For details concerning the interpretation of 
our data we refer to \citet[][Paper~I]{RoDe01}.

Evidence for the disk--halo interaction has also been gathered from 
observations tracing other constituents of the ISM, such as the hot ionized 
medium (HIM) \citep{Br94,Da97}, and dust \citep{How99}. The latter 
constituent is also briefly discussed for our survey galaxies in the 
following section. 

\section{Extraplanar dust}

As a by--product of our continuum subtraction process the R--band images 
were also used to study another important constituent of the ISM, 
namely the dust. Edge--on galaxies are ideally suited candidates for this 
kind of investigation, since the dust lane is clearly visible in projection, 
although it should be noted that not all edge--on galaxies bear a bright and 
filamentary dust lane. In recent studies of the edge--on spiral NGC\,891, 
using high spatial resolution broad band imaging in various filterbands, 
high latitude dust features have been discovered \citep{How97,Ro01}. 
Hundreds of dust features are recognized from which over a dozen individual 
dust features have already been studied in more detail \citep{How97}. They 
are located at distances of 300\,pc $\leq z \leq$ 1500\,pc above the 
galactic plane. Although other investigations dealing with dust in edge--on 
galaxies have been carried out in the past \citep[e.g.,\,][]{So87,So94}, the 
recent works by \citet{How97,How99} are one of the first linking the 
extraplanar dust features to the interaction between the disk and the halo 
of the galaxy, although \citet{So87} already argued that the dust features 
might be connected to a magnetic process termed {\em magnetic fountain}.

\subsection{Analysis}

In order to study the individual dust features in more detail the R--band 
images have been processed to enhance the visibility of the dust structures 
against the galaxy background. As performed in the studies by 
\citet{How97,How99}, we have similarly applied an unsharp--masking method to 
our R--band images. We have gaussian--filtered the images with a FWHM of 
10--15\,pixels for the different data sets. Then the original R--band frames 
have been divided by the gaussian-filtered images to produce the final 
unsharp--masked images. By applying this procedure to the whole frame, point 
sources with high count--rates (i.e.~bright stars) appear as artifacts. 

The detection of individual dust features is a straightforward method, since
in the positive grey scale images the white structures (i.e.~the dust) 
clearly separated from other morphological features which are recognizable in 
the galaxy background (stars, remaining cosmics, gas, background galaxies). 
This method was only applied to enhance the visibility. Possible artifacts, 
which arise due to the smoothing process, can easily be identified on the 
original R--band frames. However, the detection of extraplanar dust features 
rests on several galaxy parameters. The inclination is a very important 
factor. Galaxies with inclinations $i\leq80^\circ$ do not separate the 
prominent dust lane clearly enough in projection. The distribution of dust 
filaments along the dust lane can be quite different for each galaxy, which 
is partly due to a projection effect, and partially attributed to the 
intrinsic morphology and distribution of the dust within each galaxy.    

\vspace{0cm}
\begin{figure}[]
\begin{center}
%\rotatebox{0}{\resizebox{8.8cm}{!}{\includegraphics{h4307newF17.ps}}}
\caption[]{\small Unsharp--masked R--band image of IC\,2135, showing very 
patchy dust filaments. No continuous dust lane is visible. N is on top, and 
E is to the left.}
\end{center}
\end{figure}
\vspace{0cm}

\vspace{0cm}
\begin{figure}[]
\begin{center}
%\rotatebox{0}{\resizebox{8.8cm}{!}{\includegraphics{h4307F18.ps}}}
\caption[]{\small Unsharp--masked R--band image of NGC\,4302, showing 
spectacular dust features above/below the galactic plane. The face--on 
spiral to the right is NGC\,4298. N is on top, and E is to the left.}
\end{center}
\end{figure}
\vspace{0cm}

\vspace{0cm}
\begin{figure}[]
\begin{center}
%\rotatebox{0}{\resizebox{8.8cm}{!}{\includegraphics{h4307newF19.ps}}}
\caption[]{\small Unsharp--masked R--band image of NGC\,4402. Several 
dust filaments, which reach high galactic latitudes, are visible. N is on 
top, and E is to the left.}
\end{center}
\end{figure}
\vspace{0cm}   

\vspace{1.0cm}  
\begin{figure*}[h]
\hspace*{3.0cm}
\hspace{-1.8cm}
%\rotatebox{0}{\resizebox{16.8cm}{!}{\includegraphics{h4307newF20.ps}}}
\caption[Unsharp--masked R--band images: \newline NGC\,3044, IC\,2531, 
NGC\,4634, NGC\,5170, IC\,4351, UGC\,10288]{\small Unsharp--masked R--band 
images. Upper left: NGC\,3044, upper right: IC\,2531, middle left: NGC\,4634, 
middle right: NGC\,5170, lower left: IC\,4351, lower right: UGC\,10288. 
The corresponding H$\alpha$ images can be found in \citet{Ro00}. Orientation: 
N is to the top and E to the left.}  
\end{figure*}
\vspace{0.2cm}

\subsection{Results}

The unsharp--masked greyscale images of the survey galaxies are shown in 
Figs.~17--20 \citep[the galaxies from our first sub--sample][]{Ro00} and 
all others in Figs.~22--54 (only available electronically at EDP Sciences) 
in the central rows (between the R--band and H$\alpha$ images) in order of 
their increasing R.A.. Each figure consists of $2\times3$ sub--panels, with 
two galaxies, each one accompanying one column. The scale is marked by a 
black horizontal bar in the lower corner of each figure, and the orientation 
is N to the top, and E to the left.

The following results are obtained from the simple analysis of the 
unsharp--masked R--band images. We find extraplanar dust (eDust) filaments 
in 26 of our 74 galaxies. 48 galaxies lack in showing extraplanar dust. In 
one case (ESO\,274--1) we cannot state with confidence whether there are 
eDust filaments visible or not, as this galaxy is superimposed onto a crowded 
field of foreground stars. The unsharp--masked process caused many artifacts, 
therefore no clear statement can be given in this particular case. We count 
this galaxy as a negative detection in our statistics, which is summarized in 
Table~4. Typical distances of the extraplanar dusty filaments from the 
galactic midplane are $|z|\sim0.6-1.5$\,kpc. 

\begin{table}[h]
\setcounter{table}{3}
\caption[]{eDust detections / non--detections}
\begin{flushleft}
\begin{center}
\begin{tabular}{cccc}
\noalign{\smallskip}
\hline\hline
eDust & no eDust & pos. corr. & neg. corr.  \\ 
\hline
n=26 & n=48 & n=66 & n=8 \\
35\% & 65\% & 89\% & 11\% \\
%\noalign{\smallskip}
\hline
\end{tabular}
\end{center}
\end{flushleft}
\end{table}

The dust structures at high galactic latitudes are diverse in morphology. 
In \object{IC\,2135}\footnote{IC\,2135 was inadvertently identified with 
NGC\,1963 in \citet{Ro00}, due to a general confusion in the literature and 
in electronic databases that existed back then.}, where no prominent dust 
lane is detected, only patchy features are seen in the disk--halo interface 
on both sides of the disk (cf. Fig.~17). Very spectacular filamentary 
structures are discovered in NGC\,4302 (Fig.~18). Many of the discovered 
dusty filaments are strongly bended, suggesting that magnetic fields may act 
upon the charged dust particles. The {\em Virgo Cluster} spiral NGC\,4402 
(see Fig.~19) shows strongly winded filaments, which are detectable out to 
$|z|\sim1.7$\,kpc. The dusty structures show a disturbed morphology as well 
as the eDIG morphology. This suggests that ram--pressure stripping, caused 
by the interaction of the ISM with the ambient {\em Virgo Cluster} 
Intracluster Medium (ICM), is acting upon NGC\,4402. This has been concluded  
for other {\em Virgo Cluster} galaxies as well \citep[e.g.,\,][]{Ve99,Vo00}. 
In Fig.~20 we show the remaining six of the nine galaxies studied in 
H$\alpha$ by \citet{Ro00}.

Candidate galaxies with prominent dusty features at high galactic latitudes 
include two starburst galaxies NGC\,3628, NGC\,5775, and also the galaxy 
NGC\,7090. The unsharp--masked images can be seen in the Figs.~35, 41, 48, 
respectively. 

The distribution of the high--$|z|$ dust features reveals that {\em usually 
extraplanar dust is visible in those galaxies, where also eDIG is detected}. 
The dust features, however, reach much lower galactic latitudes, except in 
two cases -- NGC\,360, NGC\,4302 -- where dust is detected at larger 
distances from the galactic midplane than eDIG. Generally, the distribution 
of the high--$|z|$ dust is restricted to $|z|\leq1.5$\,kpc.    

We find two cases, where eDust was detected (NGC\,360, ESO\,240--11), but 
where no eDIG has been detected. On the other side, we find six cases where 
the opposite is true. The majority (89\,\%), however, shows a clear 
correlation between eDIG/eDust detections and non--detections. Although a 
correlation of the presence/non--presence of high--$|z|$ gas and dust exists, 
it should be noted that {\em in general} no 1:1 correlation exists (i.e. 
individual gas and dust filaments are not spatially correlated)! This might 
indicate that other mechanisms are responsible for the transport of the dust 
into the halo, than for the gas transport.  

\subsection{Discussion}

The presented results of this investigation of the extraplanar dust 
distribution in edge--on galaxies are in very good agreement with those by 
\citet{How99,How00}. They investigated a small sample of 12 nearby 
edge--on spirals using multicolor broad band (B and V) observations 
\citep{How99}. Four of their studied objects are also covered in our 
H$\alpha$ survey, namely NGC\,891, NGC\,3628, NGC\,4302, and NGC\,4634. 
Based on their small sample they derive physical properties for a few 
individual dust features. They derive large column densities of 
$3\times10^{20}\,\rm{cm^{-3}} \leq N_{\rm{H}} \leq 2\times10^{21}\,
\rm{cm^{-3}}$, and derive dust masses of $\sim$ a few $10^{5}\,
\rm{M_{\odot}}$. Although their observations have been carried out in the 
B, and V bands, and our observations were performed in the somewhat more 
transparent R--band, we find a good general agreement between their and 
our data for the structure of the high--$|z|$ dust features. 

The coincidence of high latitude gas and dust structures suggests that these 
both phases of the ISM are most likely tied to the same driving force, the 
star formation processes in the underlying galaxy disk. However, the 
mechanism for the dust ejection might be different from the ones responsible 
for expelling the gas to high galactic latitudes. While models for the gas 
transport include magneto--hydrodynamic (MHD) flows such as the {\em galactic 
fountains} \citep{Sh76}, {\em chimneys} \citep{No89}, favored expulsion 
models for the dust may be considered via radiation pressure on dust grains, 
termed {\em photolevitation} \citep[e.g.,\,][]{Fr91,Fe91}, and flows 
initiated by magnet field instabilities ({\em Parker instabilities}), caused 
by SNe explosions \citep[e.g.,\,][]{Pa92,Sh01}.    

The reason to assume that the two distinct phases of the ISM are expelled 
by different processes is justified by the basic notion, that dust grains in 
the harsh environments of the star formation processes will be completely 
destroyed, and should not be observed at high galactic latitudes.

\section{Summary and conclusions}

The detection of star formation driven gaseous outflows using the H$\alpha$ 
narrow band line imaging techniques is a viable method to trace the 
distribution of the warm ionized medium in external galaxies on a global 
scale. Many of the actively star-forming galaxies show similar, yet 
different, morphologies as the starburst galaxies. 

We have presented the individual results for the H$\alpha$ survey galaxies. 
From the 74 investigated edge--on spirals we have detected eDIG in 30 
galaxies, that is 40.5\%. We can therefore conclude, that {\em the presence 
of eDIG} in halos of galaxies is not a unique case for only a few galaxies, 
rather it {\em is found to be ubiquitous in galaxies, which exceed a certain 
level of SFR per unit area}, or even at a fainter threshold in combination 
with enhanced dust temperatures \citet{RoDe01}. However, it can thus also be 
concluded that eDIG is not a common feature among {\em all} late--type 
spiral galaxies, as many of them do not show eDIG (at the level of the 
observed sensitivities). The presence of eDIG is depending on the SF 
activity on both local and global scales. 

The morphology of eDIG shows a wide variety ranging from individual plumes, 
and filaments in galaxies with mediocre SF activity, to pervasive layers in 
the actively SF galaxies. A few of our eDIG detected targets bear a more or 
less intense layer of extended emission with typical extraplanar distances 
of 1.5--2\,kpc. Individual filaments of some galaxies (e.g.,\,NGC\,4388, 
NGC\,5775) even reach distances of up to $\sim6$\,kpc. In the case of 
NGC\,4700 a good correlation between extended H$\alpha$ emission and radio 
continuum (radio halo) is found, which further strengthens the disk--halo 
interaction scenario.  

%__________________________________________________________________

\begin{acknowledgements}
It is our sincere pleasure to express our thanks to Dr.~Francisco Prada for 
carrying out some of the observations at Calar Alto in an emergency case. 
We owe special thanks to Dr.~Michael Dahlem for providing us with the data on 
NGC\,3936, kindly observed by Dr.~Eva Grebel. We would also like to thank 
the anonymous referee for his/her helpful comments. The authors would like 
to thank Deutsches Zentrum f\"ur Luft-- und Raumfahrt (DLR) for financial 
support of this research project through grant 50\,OR\,9707. Additional 
travel support for the Calar Alto observing runs is acknowledged from the 
DFG through various grants. This research has made extensive use of the 
NASA/IPAC Extragalactic Database (NED) which is operated by the Jet 
Propulsion Laboratory, California Institute of Technology, under contract 
with the National Aeronautics and Space Administration. The Southern 
H$\alpha$ Sky Survey Atlas (SHASSA) is supported by the National Science 
Foundation.  
\end{acknowledgements}

\clearpage

\begin{table*}[h]
\setcounter{table}{4}
\caption[DIG morphology of the survey galaxies]{DIG morphology of the survey 
galaxies}
\begin{flushleft}
\begin{center}
\begin{tabular}{lcccc}
\noalign{\smallskip}
\hline\hline
Galaxy & DIG morph.$^{\footnotesize a}$ & vertical extent & radial extent & 
Notes \\ 
& & $|z|$\,[kpc] & R$_{\rm SF}$\,[kpc] & \\ \hline
NGC\,24 & d, n & 0.68 & ~\,3.09 & not perfectly inclined\\
NGC\,100 & n & 0.63 & ~\,5.82 & \hbox{H\,{\sc ii}} regions in the disk\\
UGC\,260 & eh2, f, h$_{\rm f}$, pec & 2.00 & ~\,7.14 & pec = tidal debris?, 
galaxy?\\
ESO\,540--16 & d & 0.73 & ~\,7.88 & strong asymmetry of planar DIG\\
MCG-2-3-16 & a, n & 0.43 & ~\,2.67 & disturbed disk\\
NGC\,360 & d & 0.59 & 11.71 & \\
NGC\,669 & n & & 28.33 & \\
UGC\,1281 & n & 0.42 & ~\,3.34 & \hbox{H\,{\sc ii}} regions not aligned\\
NGC\,891 & ee, eh2, f, h$_{\rm b}$ & 2.15 & ~\,9.85 & eDIG asymmetry 
(north--south) \\
UGC\,2082 & d, n & 0.44 & ~\,4.68 & \\
IC\,1862 & d, n & & 47.17 & slightly warped disk\\
NGC\,1247 & d, pa & & 36.10 & \\
ESO\,117--19 & d & & 27.89 & \\
IC\,2058 & n & 0.48 & ~\,7.66 & \\
ESO\,362--11 & h$_{\rm f}$ & 2.47 & ~\,7.80 & \\
ESO\,121--6 & h$_{\rm f}$ & 1.42 & ~\,7.88 & \\
NGC\,2188 & ee, eh2, f, pl & 1.35 & ~\,6.18 & \\
ESO\,209--9 & h$_{\rm b}$, pa, pl & 1.57 & 12.60 & + galactic emission \\
UGC\,4559 & n & 1.00 & ~\,6.29 & disk em. restr. near nucleus\\
NGC\,2654 & d, n & 1.18 & ~\,7.18 & clustered \hbox{H\,{\sc ii}} regions\\
NGC\,2683 & d, n & 0.87 & ~\,2.63 & strong disk emission\\
NGC\,3003 & d, n & & ~\,9.35 & galaxy not perfectly edge--on\\
NGC\,3221 & ee, h$_{\rm f}$ & 3.82 & 35.27 & \\
NGC\,3365 & d, n & 1.00 & ~\,2.23 & strong local disk emission\\
NGC\,3501 & n & 0.85 & ~\,7.19 & disk \hbox{H\,{\sc ii}} regions not aligned\\
NGC\,3600 & ee & 1.24 & ~\,4.20 & nuclear outflow?, warped disk\\
NGC\,3628 & ee, eh2, f & 3.13 & & prominent nuclear outflow\\
NGC\,3877 & ee, f, pl & 1.38 & 12.00 & clustered DIG emission\\
NGC\,3936 & n & & 11.46 & \\
ESO\,379--6 & ee, pl & 2.08 & 25.23 & \\
NGC\,4206 & d, pa & 1.22 & ~\,8.02 & \\
NGC\,4216 & d, n & & 11.48 & strong H$\alpha$ bulge emission \\
NGC\,4235 & ee, h$_{\rm f}$ & 2.50 & ~\,4.58 & dust obscur. near nucleus \\
NGC\,4256 & d, n & & 13.17 & planar DIG asymmetry\\
NGC\,4388 & f, h$_{\rm f}$, pa & 5.92 & 15.50 & prominent halo patch\\
NGC\,4700 & h$_{\rm b}$, f, pa & 2.18 & 12.50 & one filament 
$z\approx3$\,kpc\\
NGC\,4945 & f, h$_{\rm b}$, pl & 5.28 & \hspace*{-0.39cm} $\geq$22.42 & 
outflow cone\\
NGC\,5290 & ee, f, h$_{\rm b}$ & 4.00 & 13.00 & nucl. outflow, starburst?\\
NGC\,5297 & d, n & & 16.00 & not perfectly edge--on \\
NGC\,5775 & h$_{\rm b}$, f, pl & 5.38 & 21.44 & prominent filaments\\
ESO\,274--1 & ee, f & 0.75 & & strong local eDIG\\
NGC\,5965 & d, n & & 19.78 & \\
NGC\,6722 & n & & 42.86 & warped disk\\
IC\,4837A & d & & 13.57 & strong local DIG\\
ESO\,142--19 & n & & & prominent dust lane\\
IC\,4872 & n & 0.71 & ~\,3.33 & \\
NGC\,6875A & ee?, pa & 2.75 & 12.00 & not perfectly edge--on\\
MCG-01-53-012 & n & 1.40 & 10.80 & \\
IC\,5052 & a, ee, h$_{\rm b}$, pl & 1.24 & ~\,9.90 & DIG distr. 
asymmetrically\\
IC\,5071 & d & & 20.85 & clustered DIG emission\\
IC\,5096 & n & 1.00 & 19.61 & strong bulge emission\\
NGC\,7064 & ee, h$_{\rm f}$, pl & 0.92 & ~\,7.20 & disk bi--sected\\
NGC\,7090 & a, ee, f, h$_{\rm f}$ & 1.78 & ~\,8.39 & strong local emission\\
UGC\,11841 & n & & & gal. barely vis. in H$\alpha$\\
NGC\,7184 & d, n & & 20.44 & strong emission in annulus\\
%\noalign{\smallskip}
\hline
\end{tabular}
\end{center}
\end{flushleft}    
\end{table*} 

\clearpage

\begin{table*}[h]
\setcounter{table}{4}
\caption[continued]{continued}
\begin{flushleft}
\begin{center}
\begin{tabular}{lcccc}
\noalign{\smallskip}
\hline\hline
Galaxy & DIG morph.$^{\footnotesize a}$ & vertical extent & radial extent & 
Notes \\ 
& & $|z|$\,[kpc] & R$_{\rm SF}$\,[kpc] & \\ \hline
IC\,5171 & d & & 10.19 & strong planar DIG\\
IC\,5176 & ee, h$_{\rm f}$ & 2.35 & 10.22 & \\
NGC\,7339 & n & 0.36 & ~\,2.32 & disk \hbox{H\,{\sc ii}} regions not aligned\\
NGC\,7361 & d, n & & ~\,9.49 & strong disk emission\\
NGC\,7412A & a, n & 0.47 & ~\,4.77 & bright disk \hbox{H\,{\sc ii}} regions\\
UGC\,12281 & n & 1.36 & 20.00 & \\
NGC\,7462 & f, h$_{\rm f}$, pl & 1.76 & 10.81 & slight asymmetry\\
UGC\,12423 & n & & & gal. barely vis. in H$\alpha$\\
NGC\,7640 & d, n & & ~\,7.11 & strong planar DIG\\
ESO\,240--11 & n & & 29.93 & slight disk asymmetry \\
%\noalign{\smallskip}
\hline
\end{tabular}
\end{center}
\hspace{2.5cm}{\footnotesize $^{a}$ a$=$arc(s), $d$=disk emission (only 
planar DIG), ee$=$extended emission (locally),\\ 
\hspace{2.67cm} eh2$=$extraplanar \hbox{H\,{\sc ii}} region(s), 
f$=$filament(s), h$_{\rm b}$=bright halo, h$_{\rm f}$=faint halo,\\
\hspace{2.67cm} n$=$no (e)DIG, pa$=$patch(es), pec$=$peculiar, pl$=$plume(s)}
\end{flushleft}    
\end{table*} 

\vspace{1.5cm}

\begin{figure*}[]
\begin{center}
\hspace*{0.9cm}
\rotatebox{0}{\resizebox{16.0cm}{!}{\includegraphics{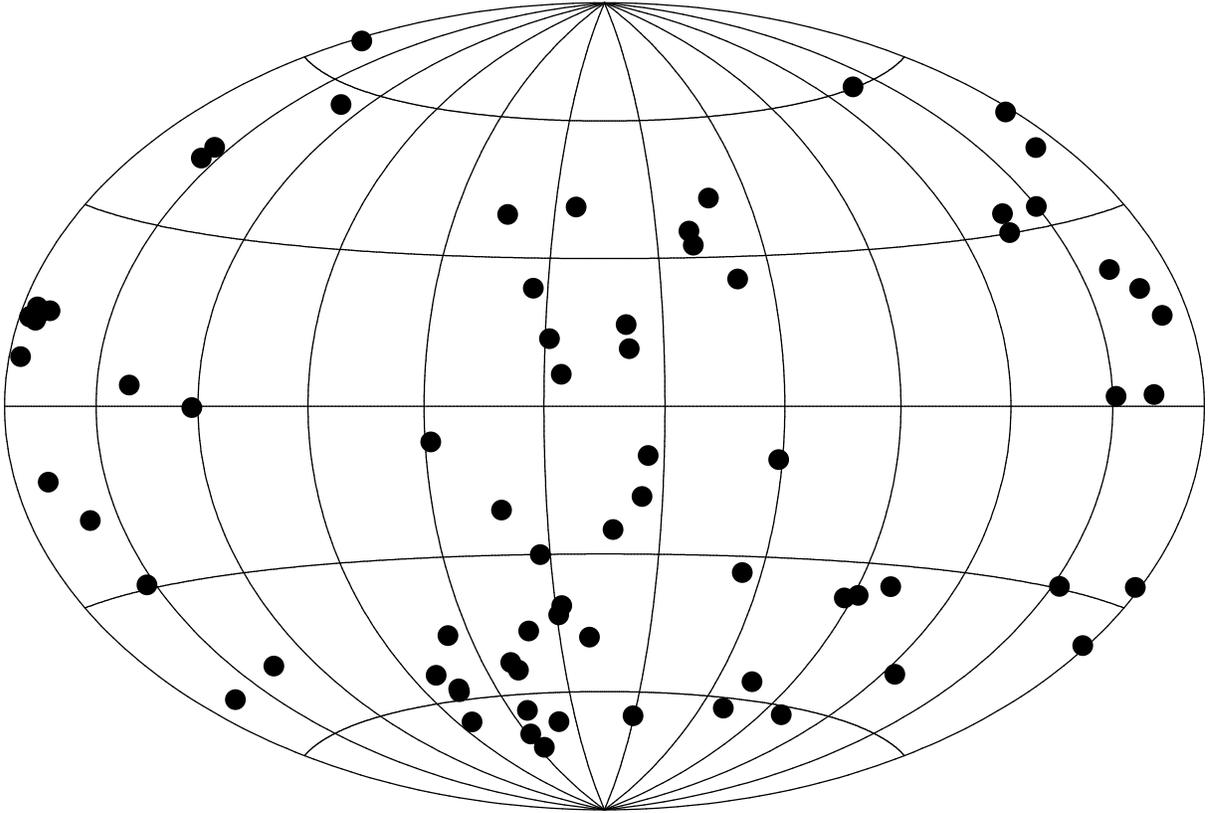}}}
\caption[]{Distribution of the observed 74 galaxies of the H$\alpha$ survey 
on the celestial sphere in Aitoff projection. The center position is 0,0 in 
R.A., Dec.}
\end{center}
\end{figure*}

\clearpage

%__________________________________________________________________

\end{document}